\documentclass{emulateapj}

\def \nh {N${\rm _H}$}

\def \arcmin {\hbox{$^\prime$}}
\def \arcsec {\hbox{$^{\prime\prime}$}}

\def\spose#1{\hbox to 0pt{#1\hss}}
\def\ltsim{$\mathrel{\spose{\lower 3pt\hbox{$\sim$}}
        \raise 2.0pt\hbox{$<$}}$\thinspace}
\def\gtsim{$\mathrel{\spose{\lower 3pt\hbox{$\sim$}}
        \raise 2.0pt\hbox{$>$}}$\thinspace}
\def \msun {${\rm M_\odot}$}
\def \mstar{${\rm M_\star}$}
\def \rvir{{\hbox{$r_{\rm vir}$}}}

\def \cvir{{\hbox{$c_{\rm vir}$}}}
\def \mvir{{\hbox{$M_{\rm vir}$}}}

\def \lcdm{{\hbox{$\Lambda$CDM}}}

\def \nh {$N_{\rm H}$}

\def \dtwentyfive {${\rm D_{25}}$}

\newcommand{\sersic}{S\'{e}rsic}
\newcommand{\apec}{APEC}

\newcommand{\rhog}{${\rho_g}$}
\newcommand{\zfe }{${\rm Z_{Fe}}$}

\newcommand{\emiss } {$\Lambda(T,Z_{\rm Fe})$}
\newcommand{\chandra }{{\em Chandra}}
\newcommand{\xspec }{{\em Xspec}}

\newcommand{\ciao }{{\em CIAO}}
\newcommand{\caldb }{{\em Caldb}}
\newcommand{\heasoft }{{\em Heasoft}}

\newcommand{\xmm }{{\em XMM}}
\newcommand{\asca }{{\em ASCA}}
\newcommand{\rosat }{{\em ROSAT}}

\newcommand\lk{\hbox{{$L_{\rm K}$}}}
\newcommand\lb{\hbox{{$L_{\rm B}$}}}
\newcommand{\lsun }{${\rm L_\odot}$}
\newcommand{\leda}{{\em{LEDA}}}
\newcommand{\ned}{{\em{NED}}}

\newcommand{\twomass}{{\em 2MASS}}
\def \dtwentyfive {${\rm D_{25}}$}

\newcommand{\thin}{\thinspace}
\newcommand\omegam{\hbox{{$\Omega_{\rm m}$}}}
\newcommand\omegalambda{\hbox{{$\Omega_{\Lambda}$}}}
\newcommand\kmsmpc{{\rm km s$^{-1}$ Mpc$^{-1}$}}
\newcommand\ho{\hbox{{$H_{0}$}}}
\slugcomment{Submitted to the Astrophysical Journal}
\shorttitle{Dark Matter and Gas Fraction in Galaxy Groups.}
\shortauthors{Gastaldello et~al.}

\begin{document}

\title{Probing the Dark Matter and Gas Fraction in Relaxed Galaxy
Groups with X-ray observations from \chandra\ and \xmm}
\author {Fabio Gastaldello\altaffilmark{1}, David A. Buote\altaffilmark{1}, 
Philip J. Humphrey\altaffilmark{1}, Luca Zappacosta\altaffilmark{1},
James S. Bullock\altaffilmark{1}, Fabrizio
Brighenti\altaffilmark{2,3}, \& William G. Mathews\altaffilmark{2}}
\altaffiltext{1}{Department of Physics and Astronomy, University of 
California at Irvine, 4129 
Frederick Reines Hall, Irvine, CA 92697-4575}
\altaffiltext{2}{UCO/Lick Observatory, University of California at Santa Cruz,
 1156 High Street, Santa Cruz, CA 95064}
\altaffiltext{3}{Dipartimento di Astronomia, Universit\`a di Bologna, via 
Ranzani 1, Bologna 40127, Italy}
\begin{abstract}
We present radial mass profiles within $\sim0.3$ \rvir\ for 16 relaxed
galaxy groups-poor clusters (kT range 1-3 keV) selected for optimal
mass constraints from the \chandra\ and \xmm\ data archives. After
accounting for the mass of hot gas, the resulting mass profiles are
described well by a two-component model consisting of dark matter
(DM), represented by an NFW model, and stars from the central
galaxy. The stellar component is required only for 8 systems, for
which reasonable stellar mass-to-light ratios ($\rm{M/L_{K}}$) are
obtained, assuming a Kroupa IMF. Modifying the NFW dark matter halo by
adiabatic contraction does not improve the fit and yields
systematically lower $\rm{M/L_{K}}$. In contrast to previous results
for massive clusters, we find that the NFW concentration parameter
(\cvir) for groups decreases with increasing \mvir\ and is
inconsistent with no variation at the $3\sigma$ level.  The
normalization and slope of the \cvir\--\mvir\ relation are consistent
with the standard $\Lambda$CDM cosmological model with
$\sigma_{8}=0.9$ (considering a 10\% bias for early forming
systems). The small intrinsic scatter measured about the
\cvir\--\mvir\ relation implies the groups represent preferentially
relaxed, early forming systems. The mean gas fraction ($f=0.05\pm
0.01$) of the groups measured within an over-density $\Delta=2500$ is
lower than for hot, massive clusters, but the fractional scatter
($\sigma_f/f=0.2$) for groups is larger, implying a greater impact of
feedback processes on groups, as expected. 
\end{abstract}

\keywords{Cosmology: observations--- dark matter--- galaxies: halos--- X-rays: 
galaxies: clusters---methods: data analysis}

\section{Introduction}

The properties of dark matter (DM) halos are a powerful discriminator
between different cosmological sce\-na\-rios of structure formation.
Dissipationless simulations of Cold Dark Matter (CDM) models find that
the radial density profiles of DM halos are fairly well described
between approximately 0.01-1 \rvir\ (where \rvir\ is the virial
radius) by the 2-parameter NFW model suggested by
\citet{nfw}.  Of particular importance is the distribution of halo
concentration (\cvir, the ratio between \rvir\ and the characteristic
radius of the density profile, $r_s$) and
\mvir, the virial mass. Low mass halos are more concentrated because
they collapse earlier than halos of larger mass, thus producing a
predicted correlation between \cvir\ and \mvir. A significant scatter
at fixed virial mass is expected and thought to be related to the
distribution of halo formation epoch
\citep[e.g.,][]{bullock01,wechsler02}. The \cvir-\mvir\ relation and
its scatter is a source of deviation from the self-similar scaling
relation expected if the observable properties of halos are driven by
simple gravitational collapse of the dominant dark matter component
\citep[e.g.,][]{thomas01}. For the currently favored $\Lambda$CDM
model the median \cvir\ varies slowly over a factor of 100 in \mvir,
whereas the scatter remains very nearly constant
\citep[e.g.,][]{bullock01,dolag04,kuhlen05}.  The precise relation
between \cvir\ and \mvir\ is expected to vary significantly as a
function of the cosmological parameters, including $\sigma_8$ and $w$,
the dark energy equation of state
\citep{dolag04,kuhlen05}, making an observational test of this relation a 
very powerful tool for cosmology.

High quality X-ray data from \chandra\ and \xmm\ observations indicate that
the NFW model is a remarkably good description of the mass profiles of
massive galaxy clusters out to large portions of their virial radii
\citep[e.g.,][]{pratt02,lewi03a,pointe05,vikh06a,zappacosta06}.
Typical values and scatter of concentrations determined from the
samples of clusters analyzed in \citet{pointe05} and \citet{vikh06a}
are in general agreement with the simulation results.  The
\cvir\--\mvir\ relation measured by \citet{pointe05}, when fitted with
a power law, has a slope of $\alpha = -0.04\pm0.03$. This slope is
quite consistent with a constant value and is marginally consistent
($\approx 2\sigma$) with $\Lambda$CDM \citep{dolag04}. The optical
study by \citet{lokas06b} using galaxy kinematics for six nearby
relaxed Abell clusters obtained results consistent with the above
X-ray studies but with larger uncertainty (e.g., $\alpha =
-0.6\pm1.3$).

Observational tests of \lcdm\ have proven controversial at the galaxy
scale \citep[see discussion in][]{humphrey06a}. Recently, using high
quality X-ray \chandra\ data, in \citet{humphrey06a} we obtained
accurate mass profiles for 7 elliptical galaxies, well described by a
two-component model comprising an NFW DM halo and a stellar mass
component. Omitting the latter component, which dominates the mass
budget in the inner regions, leads to unphysically large
concentrations \citep[see also][]{mamon05a} and may explain some large
values found in the literature for elliptical galaxies
\citep{sato00,khosroshahi04}.  The measured \cvir-\mvir\ relation of
the 7 galaxies generally agrees with $\Lambda$CDM, provided the
galaxies represent preferentially relaxed, earlier forming systems.

Very few constraints exist on the group scale, where the simulations
of DM halos are more reliable, compared to massive clusters, because a large 
number of objects can be simulated at once \citep[e.g.,][]{bullock01}.
\citet{sato00} investigated the $c$-$M$ relation in X-rays using \asca\ 
for a sample of objects ranging from $10^{12}$ to $10^{15}$~\msun,
including objects in the mass range discussed in this paper. (However,
neither the names of the objects in their sample, nor the description
of the data reduction and analysis, has appeared in the literature.)
The slope obtained for the $c$-$M$ relation was steep,
$-0.44\pm0.13$. The limited spatial resolution of \asca\ and energy
dependence of its PSF made problematic the determination of reliable
density and temperature profiles, and the authors neglected any
stellar mass component in their fits.  Optical studies of groups using
galaxy-galaxy lensing \citep{mandelbaum06} and caustics in redshift
space \citet{rines06} obtain \cvir-\mvir\ relations that are
consistent with CDM simulations and with no variation in $c$ with $M$,
but with large errors.

The scale of galaxy groups is also particularly inte\-re\-sting for the
investigation of the influence of baryons on the DM profile. While the
stellar mass component is clearly distinguished from the NFW DM
component in the gravitating mass profiles obtained from \chandra\
observations of elliptical galaxies \citep{humphrey06a}, X-ray studies
of relaxed clusters do not report significant deviations from a single
NFW profile fitted to the gravitating mass
\citep{lewi03a,pointe05,vikh06a,zappacosta06}, except for a few
group-scale objects \citep{vikh06a}. The group scale seems to
represent a transition in the character of the mass profiles
\citep[and temperature profiles,][]{humphrey06a} and needs
to be systematically explored.

X-ray studies of mass profiles in galaxy systems have the advantage
that the pressure tensor of the hot gas is isotropic and the gas in
hydrostatic equilibrium (HE) traces the entire 3D cluster potential
well.  If one is careful to choose relaxed objects (i.e., with smooth,
regular X-ray images) then hydrostatic equilibrium is a good
approximation and the resulting gravitating mass is reliable, accurate
to at least $\sim 15$\% even in the presence of turbulence
\citep[e.g.,][]{evrard96,faltenbacher05,nagai07}. Because of limitations of
previous X-ray telescopes like \rosat\ and \asca, some simplifying
assumptions like isothermality had to be made for the determination of
group masses \citep[see][and references therein]{mulch00}.
\chandra\ and \xmm\ have provided for the first time high quality, 
spatially resolved spectra of the diffuse hot gas of X-ray groups 
\citep[e.g.,][]{buot03a,sun03,osul03b,buot04b,pratt05}.

An investigation of the detailed mass profiles of galaxy groups
($M=10^{13}$-$10^{14}$ \msun) with higher quality \chandra\ and \xmm\
data is, therefore, timely.  In this paper we present measurements of
the mass profiles of a sample of 16 groups chosen to provide the best
mass determinations with current X-ray data.  We selected the objects
both to be the most relaxed systems (i.e., very regular X-ray image
morphology), to insure hydrostatic equilibrium is a good
approximation, and to have the highest quality \chandra\ and \xmm\
data, which allow for the most precise measurements of the gas density
and temperature profiles. This paper is part of a series
\citep[see also][]{humphrey06a,zappacosta06,buote06b} using 
high-quality \chandra\ and \xmm\ data to investigate the mass profiles of 
galaxies, groups and clusters, placing constraints upon the \cvir-\mvir\ 
relation over $\approx 2.5$ orders of magnitude in \mvir.
It is also the first in a series investigating the X-ray 
properties of groups and poor clusters: in future papers we will investigate
the entropy and heavy element abundance profiles.
This paper is organized as follows. In \S~\ref{sect_targets} we discuss the 
target selection and in \S~\ref{sect_reduction} the data-reduction.
We discuss the spectral analysis in \S~\ref{sect_spectra}, the
mass analysis in \S~\ref{sect_mass} and present the results in 
\S~\ref{sect_results}. We discuss the results for individual objects in the
sample in \S~\ref{sect_objects}, comparing with previous work in the
literature.  The systematic uncertainties in our analysis are
discussed in \S~\ref{sect_systematics}, and we present a discussion of
our results in \S~\ref{sect_discussion} with our conclusions in
\S~\ref{sect_conclusions}.  All distance-dependent quantities have
been computed assuming \ho = 70 \kmsmpc,
\omegam = 0.3 and \omegalambda = 0.7. Our assumed virial radius is defined as
the radius of a sphere of mass \mvir, the mean density of which (for
redshift 0) is 101 times the critical density of the universe
(appropriate for the assumed cosmological model) and estimated at the
redshift of the object. We will quote values for concentrations and
masses at different over-densities to ease comparison with previous
work in Appendix \ref{sect_overdensities}. Our analysis procedure is
described in greater detail in Appendix \ref{sect_deproj_equations}.
All the errors quoted are at the 68\% confidence limit.

\begin{deluxetable*}{llcclllllc}
\tablecaption{The group sample and journal of observations\label{table_obs}}
\tabletypesize{\scriptsize}
\tablehead{
\colhead{Group} & \colhead{z} & \colhead{Dist} & \colhead{ACIS} & \colhead{\chandra\ exp} & \colhead{EPIC} & \colhead{pn} & \colhead{\xmm\ exp} & $r_{out}$$^{c}$ & $\Delta^{d}$ \\
\colhead{} & \colhead{} & \colhead{Mpc} & \colhead{Aim point} & \colhead{(ks)} & \colhead{filter} & \colhead{mode} & \colhead{(ks)} & \colhead{kpc} & \colhead{} \ 
}

\startdata
NGC 5044    &  0.0090    & 38.8 & S &  20.2       & Thin    & FF  & $19.5/19.3/8.9 + 38.4/38.4/32.0^{b}$ & 326 & 101.9 \\
NGC 1550    &  0.0124    & 53.6 & I &  $9.8 + 9.6^{a}$  & Medium  & FF  & 21.4/22.6/17.8 & 213 & 102.2 \\
NGC 2563    &  0.0149    & 64.5 &   &             & Medium  & FF  & 20.4/20.8/16.5 & 219 & 102.4 \\
A 262       &  0.0163    & 70.7 & S &  28.7       & Thin    & EFF & 23.5/23.4/15.0 & 254 & 102.5 \\ 
NGC 533     &  0.0185    & 80.3 & S &  36.7       & Thin    & FF  & 38.1/37.4/30.1 & 271 & 102.7 \\  
MKW 4       &  0.0200    & 87.0 & S &  29.8       & Medium  & EFF & 14.0/13.9/9.4  & 336 & 103.1 \\   
IC 1860     &  0.0223    & 97.1 &   &             & Thin    & FF  & 34.1/34.8/28.0 & 323 & 103.1 \\
NGC 5129    &  0.0230    & 100.2 &  &             & Medium  & FF  & 10.9/12.0/10.7 & 241 & 103.1 \\
NGC 4325    &  0.0257    & 112.2 & S &  30.0       & Thin    & FF  & 20.8/20.8/14.7 & 238 & 103.3 \\
ESO 5520200 &  0.0314    & 137.7 &  &       & Thin    & EFF & 32.2/32.2/26.7 & 418 & 103.8 \\
AWM 4       &  0.0317    & 139.0 &  &             & Medium  & EFF & 17.5/17.2/12.5 & 455 & 103.9 \\
ESO 3060170 &  0.0358    & 157.5 & I &  $13.8 + 13.9^{a}$ &        &               & & 245 & 104.2    \\ 
RGH 80      &  0.0379    & 167.0 & S &  38.5       & Thin    & EFF & 32.8/32.6/26.3 & 533 & 104.4 \\
MS 0116.3-0115 &  0.0452 & 200.2 & S &  39.0       &         &     &       & 350 & 105.0 \\  
A 2717      &  0.049     & 217.7 &   &                & Thin    & FF  & 49.2/49.6/42.9 & 730 & 105.3 \\       
RXJ 1159.8+5531 &  0.081 & 368.0 & S &   75.0       &         &     &        & 625  & 108.0 \\ 
\enddata
\tablecomments{Listed above are the groups in our sample.
Redshifts were obtained from \ned\ and the distance is the inferred luminosity 
distance for a 
cosmological model with \ho = 70 \kmsmpc,\omegam = 0.3 and \omegalambda = 0.7. 
The ACIS aim-point refers to S if 
the aim-point is located on the S3 chip or to I if the aim-point is located 
on one of the ACIS-I chips. The ACIS mode of all the observations was the Very Faint mode. 
The pn mode refers to FF if it is Full Frame or to EFF if it is Extended Full Frame; 
the MOS detectors were always in FF mode. 
The exposure times are net exposure times, after flare cleaning as described 
in the text, and for \xmm\ they refer to MOS1/MOS2/pn.\\
$^{a}$ observed twice with ACIS-I, with the
core centered on ACIS-I0 in one occasion and on ACIS-I1 in the other.\\
$^{b}$ the first set of exposures refers to the central pointing and the
second set to the offset pointing.\\
$^{c}$ the outer radius used in our analysis.\\
$^{d}$ the assumed over-density, calculated at the redshift of the object.
}
\end{deluxetable*}

\section{Target Selection}\label{sect_targets}

For this study we choose, whenever possible, to focus on X-ray bright 
objects observed by both \chandra\ and \xmm\ to exploit the complementary characteristics 
of the two satellites. The unprecedented 
spatial resolution of \chandra\ allows the temperature and density 
profiles to be resolved in the core, allowing us to disentangle the stellar 
and dark matter components. The unprecedented 
sensitivity of \xmm\ ensures good S/N even in the faint outer regions, which 
is crucial because good constraints on the virial mass of the halo require 
density and temperature constraints over as large a radial range as possible.
We looked for bright objects in the temperature range 1-3 keV with sufficiently
long exposures in the \chandra\ and \xmm\ archives, together with  
our proprietary data. The potential targets were processed 
(\S~\ref{sect_reduction}) and the images in the 0.5-10 keV band 
examined (\S~\ref{sect_images}) for evidence of disturbances:
we choose objects which have a very regular X-ray morphology, showing no or only weak signs 
of dynamical activity, with the peak of the emission coincident 
with a luminous elliptical galaxy which is the most luminous group member. 
The only exception is RGH80 which has two elliptical galaxies 
of comparable sizes in the core and probably a submerging group in the south
\citep{mahdavi05}. We include this object because it is part of a complete, X-ray
flux--limited sample of 15 groups that is scheduled to be observed by
\chandra. It also allows an interesting comparison of derived mass properties with
those obtained for the obviously relaxed systems in the sample.

The details of the observations are given in Table \ref{table_obs}. We
do not consider for analysis the available \xmm\ observations of ESO
3060170, MS 0116.3-0115 and RXJ 1159.8+5531, because they are heavily
contaminated by flares. We also do not consider the \chandra\
observation of ESO 5520200 because of insufficient S/N for our
purposes. In order to use as large a radial range as possible for
objects observed in the ACIS-S configuration but lacking \xmm\ data
(MS 0116.3-0115 and RXJ 1159.8+5531), we follow \citet{vikh06a} and
also use the ACIS-S2 CCD in the analysis.

The present sample has been selected preferentially for X-ray brightest
and most relaxed groups to obtain the best constraints on the mass
profiles in individual objects with current data. Although the sample
is biased and is not statically complete, our analysis of these
systems represents an essential step in the investigation of DM in
galaxy groups with X-rays.  In future work we will compare these
results to those obtained using the complete, X-ray flux-limited
sample of 15 groups noted above.

\begin{figure*}[t]
\vskip -0.2cm
\centerline{\includegraphics[scale=0.8]{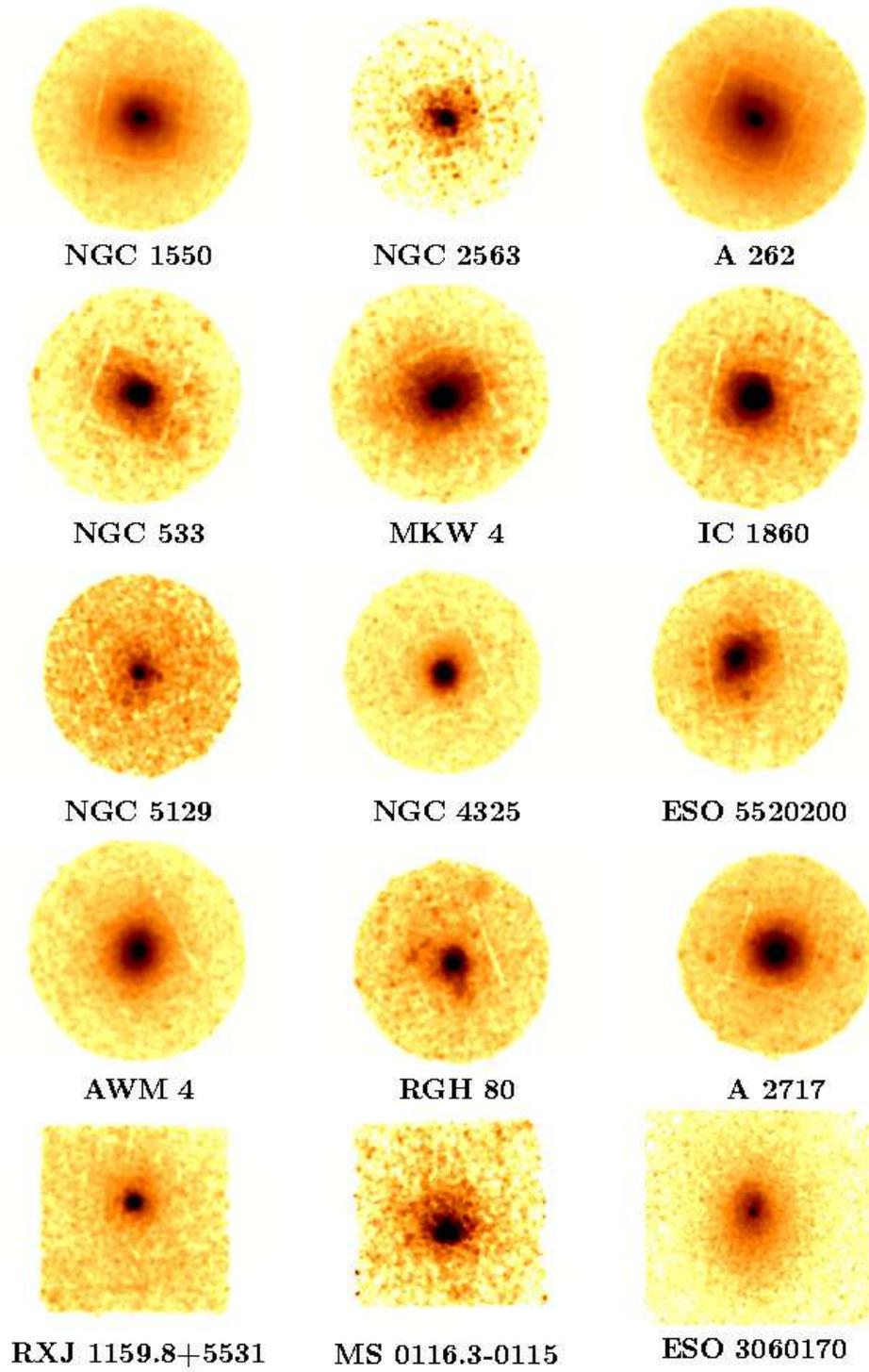}}
\caption{\label{fig.1} X-ray images of each of the objects in the
sample (see \S \ref{sect_images}).}
\end{figure*}


\section{Data reduction} \label{sect_reduction}

\subsection{Chandra}

For data reduction we used the \ciao\thin 3.2 and \heasoft\thin 5.3
software suites, in conjunction with the \chandra\ \caldb\ calibration
database 3.0.0.  In order to ensure the most up-to-date calibration,
all data were reprocessed from the ``level 1'' events files, following
the standard \chandra\ data-reduction
threads\footnote{{http://cxc.harvard.edu/ciao/threads/index.html}}.
We applied corrections to take account of a time-dependent drift in
the detector gain and, for ACIS-I observations, the effects of
``charge transfer inefficiency'', as implemented in the standard
\ciao\ tools.  From regions of least source contamination of the CCDs
we extracted a light-curve (5.0-10.0~keV) to identify periods of high
background.  Point source detection was performed using the \ciao\
tool {\tt wavdetect} \citep{freeman02}. The source lists created in
different energy bands (so as to identify unusual soft or hard
sources) were combined, and duplicated sources removed. The final list
was checked by visual inspection of the images. The resolved point
sources were finally removed so as not to contaminate the diffuse
emission. Further details about the \chandra\ data reduction can be
found in \citet{humphrey05a}.

\subsection{XMM}

We generated calibrated event files with SAS v6.0.0 using the tasks
{\em emchain} and {\em epchain}. We considered only event patterns
0-12 for MOS and 0 for pn. Bright pixels and hot columns were removed
by applying the expression (FLAG == 0) to the extraction of spectra
and images. We correct statistically for the pn out-of-time (OoT)
events. Following the standard procedure, we generate an OoT event
list, processed in the same way as the observation, and then subtract
it from the images and spectra, after being multiplied by the mode
dependent ratio of integration and read-out time (6.3\% for Full Frame
and 2.3\% for Extended Full Frame). The energy scale of the pn over
the whole spectral bandpass has been further improved using the task
{\em epreject}.  We clean the data for soft proton flares using a
threshold cut method by means of a Gaussian fit to the peak of the
histogram of the 100s time bins of the light curve \citep[see Appendix
A of][]{pratt02,deluca04} and excluding periods where the count rate
lies more than 3$\sigma$ away from the mean. The lightcurves were
extracted from regions of least source contamination (excising the
bright group core in the central 5\arcmin\ and the point source list
from the SOC pipeline, after visual inspection) in two different
energy bands: a hard band, 10-12 keV for MOS and 10-13 keV for pn, and
a wider band, 0.5-10 keV, as a safety check for possible flares with
soft spectra \citep{nevalainen05,pradas05}. The flaring periods thus
determined where further checked by visual inspection of the light curves.  
Point sources
were detected using the task {\em ewavelet} in the energy band 0.5-10
keV and checked by eye on images generated for each detector. Detected
point sources from all detectors were merged, and the events in the
corresponding regions were removed from the event list, using circular
regions of 25\arcsec radius centered at the source position. The area
lost due to point source exclusion, CCD gaps and bad pixels was
calculated using a mask image. Redistribution matrix files (RMFs) and
ancillary response files (ARFs) were generated using the SAS tasks
{\em rmfgen} and {\em arfgen}, the latter in extended source
mode. Appropriate flux-weighting was performed for RMFs, using our own
dedicated software, and for ARFs, using exposure-corrected images of
the source as detector maps (with pixel size of 1\arcmin, the minimum
scale modeled by {\em arfgen}) to sample the variation in emission,
following the prescription of \citet{saxton02}.  Spectral results
obtained using ARFs are completely consistent with the other
frequently employed method \citep[e.g.][]{arna01} of correcting
directly the spectra for vignetting \citep{gasta03,morris05}.

\subsection{Background subtraction} \label{sect_bkg}

Insuring proper background subtraction is one of the key challenges
associated with the spectral fitting of low surface brightness,
diffuse, X-ray emission. The background experienced by both \chandra\
and \xmm\ consists of (1) an extreme time variable component due to
soft (E $\sim$ tens of keV) protons channeled by the telescopes
mirrors, (2) a slowly changing (with variability time scale much
longer than the length of a typical observation) quiescent component
due to high energy particles (E $>$ a few MeV), and (3) the sky X-ray
background, decomposed into the extragalactic Cosmic X-ray background
by AGN and the Galactic X-ray emission \citep[e.g][]{lumb02,mark03a}.
 
The ``blank fields'' distributed by the observatories are not a
perfect representation of the background in any one
observation. Firstly, there are significant long term variations in
the quiescent particle background. Secondly, the soft Galactic
background component varies strongly from field to field. Finally,
there may be some residual mild flaring.

Two approaches have been investigated to obtain more accurate
background estimates than provided by the ``blank field'' background
templates: the double subtraction technique
\citep[see details in Appendix A of][]{arna02} and a complete modeling of 
the various background components
\citep[e.g.][]{lumb02,mark03a}. Double subtraction is, in principle,
very effective, but particular care has to be taken to locate a region
in the field of view of the observation completely free of source
emission; this is difficult for nearby objects. The complete modeling
of the various background components can rely on a large number of
observations performed to characterize the quiescent particle
background component (stowed or Dark Moon for \chandra, {\em closed}
for \xmm), and on the large number of observations which constitute
the ``blank field'' data sets to characterize the sky background
components. The drawback is that the resulting model, which also
includes a source component, is complicated, and parameter
degeneracies can arise. However, the method is particularly effective
for studying groups and poor clusters because the source component
(mainly characterized by the $\sim 1$~keV Fe-L shell blend) is clearly
spectrally separated from all the other background components. For the
use of this approach to the \chandra\ data we refer the reader to
\citet{humphrey05a}. Here we will describe the procedure used for
\xmm\ data which elaborates and updates the procedure described in
\citet{buot04b}.  The algorithm implemented has the following main
steps:

\begin{itemize}

\item Characterization of the quiescent particle background. We co-add
individual spectra taken from {\em closed} observations. The spectra
in the 0.4-12 keV band for MOS and 0.4-13 keV band for pn can be
adequately described by a broken power law continuum and several
Gaussians for the instrumental lines. Typical values for the model
parameters are: 0.7-0.8 for the slope at low energies, 1.0-1.2 keV for
the break energies, 0.2 for the high energy slope for MOS, 0.4-0.5 for
the high energy slope for pn. While the low-energy slope exhibits
significant variation between the observations in our study, the
high-energy slope is very stable. These results are consistent with
previous studies \citep{lumb02,deluca04,nevalainen05}.  The spectral
shape of the continuum broken power law does not change significantly
across the detector, nor does it vary in time \citep[as in the MOS
study by][]{deluca04} at high energies.

\item The model derived from the {\em closed} data is fitted to the
spectra of the Out of Field of View (OFV) events of each observation.
Portions of the MOS and pn detectors are not exposed to the sky, and
therefore neither cosmic X-ray photons nor low energy
particle--induced events (like from soft protons) are collected.
Indeed, while this is almost exactly true for the MOS (the fraction of
OoT events is 0.35\% in FF mode), for the pn a higher fraction of in
FOV events events are assigned to the OFV region as OoT (6.3\% in FF
mode and 2.3\% in EFF). In the case of strong flaring, this OoT
contribution can seriously affect the pn OFV spectrum.  We therefore
chose to extract OFV events for the pn after the flare cleaning.  The
model derived from the OFV data is taken as the initial representation
of the quiescent particle background for the particular observation.

\item To the broken power-law plus Gaussian lines describing the quiescent 
instrumental background (not vignetted and implemented as a background
model in \xspec), we add the components describing the sky X-ray
background, following \citet{lumb02}: a power law with slope $\Gamma$
= 1.41 and normalization as reported in \citet{deluca04}, free to vary
within the cosmic variance as $\Omega^{-1/2}$ \citep{barcons00} where
$\Omega$ is the solid angle covered by the observation; two thermal
components with temperatures 0.07 and 0.20~keV respectively, and
abundances fixed at solar. When modeling sources projected toward the
North Polar Spur (NGC 5129) we found it necessary to add a third
thermal component at $\sim0.4$ keV, in agreement with \citet{mark03a}
and \citet{vikh05a}.

\item This model, plus a source component described by a thermal plasma with 
temperature and abundance free to vary, was fitted jointly to the
outermost annuli used in the spectral extraction (see below). The
parameters of the source component were free to vary in each annulus.
The normalizations of the cosmic components and of the broken power
law component were scaled according to the ratio of the annuli area,
while the normalizations of the instrumental lines were free to vary,
given the fact that these components are highly spatially variable
\citep[e.g.][]{deluca04,lumb02}. An inter-calibration constant free to
vary between 0.9 and 1.1 was added to the model to take into account
any cross-calibration differences between the three EPIC
instruments. Given the best fit model for these annuli, we generate a
pulse height amplitude (PHA) correction file used in \xspec.

\item For the annuli not involved in the background fitting, we scale
the resulting model to the area of the annular region of interest in
the spectral extraction and generate a PHA file. To take into account
the variable instrumental lines, we renormalized the instrumental line
components in the model using the corresponding regions extracted from
the background templates.

\end{itemize}

We mention that possible slight variations in the particle continuum
across the detector plane \citep[see Appendix A of][]{deluca04}, or
residual mild flaring, has been modeled by slightly changing the high
energy slope of the broken power law. This does not have any tangible
effect on the spectral parameters derived for soft X-ray sources like
the objects considered in this paper.

\def\arraystretch{1.25}
\begin{deluxetable*}{lccccccc}
\tablecaption{Optical properties of the group central galaxy \label{tab:optical}}
\tablehead{
\colhead{Name\hspace*{15mm}} &
\colhead{Group} &
\colhead{$r_e$ (B)} &
\colhead{$r_e$ (K)} &
\colhead{\lb} &
\colhead{\lk} \\
 & &\colhead{kpc (arcsec)} &\colhead{kpc (arcsec)} & \colhead{$10^{10}$ \lsun} & \colhead{$10^{11}$ \lsun} & }
\startdata
NGC 5044 & NGC 5044 & \dotfill  & 4.53 (24.5) & 6.98 &  2.87  & \\
NGC 1550 & NGC 1550 & 6.45 (25.5) & 3.05 (12.1) & 4.33 & 2.09 & \\
NGC 2563 & NGC 2563 & 5.89 (19.3) & 3.73 (12.2) & 3.84 & 2.66 & \\
NGC 708 & A 262 & 25.60 (77.1) & 10.16 (30.6) & 3.84 & 4.12 & \\
NGC  533 &  NGC 533 & 16.92 (45.4) & 9.22 (25.2) & 12.4 & 6.14 & \\
NGC 4073 & MKW 4 & 19.24 (47.5) & 10.25 (25.3) & 13.7 & 7.18 & \\
IC 1860 & IC 1860 & 8.34 (18.5) & 8.03 (17.8) & 6.08 & 4.38 & \\
NGC 5129 &  NGC 5129 & 13.34 (28.7) & 6.60 (14.2) & 12.0 & 4.99 & \\
NGC 4325 & NGC 4325 &  \dotfill &  5.22 (10.1) & 4.61 & 2.33 & \\
ESO 552-020 & ESO5520200 & \dotfill & 15.7 (25.0) & 15.6 & 8.19 & \\
NGC 6051 & AWM 4 & 10.21 (16.1) & 10.33 (16.3) & 9.91 & 7.50 & \\
ESO 306-017 & ESO3060170 & \dotfill & 18.51 (26.0) & 18.5 & 6.95 & \\ 
MCG 6-29-77 & RGH80 & \dotfill & 5.11 (6.8) & \dotfill & 2.93 & \\ 
MCG 6-29-78 & RGH80 & \dotfill & 6.23 (8.3) & 4.21 & 2.39 & \\    
UGC 842 &  MS 0116.3-0115  & \dotfill  & 9.69 (10.9) & 9.29 &  5.77 & \\
ESO 349-22 & A 2717 & \dotfill  & 15.53 (16.2) & 9.20 &  5.42 & \\  
2MASSX J11595215 & RXJ 1159.8+5531 & \dotfill & 9.77 (6.4)  & 23.6 & 10.3 & \\
\enddata
\tablecomments{The optical properties of the central galaxy of each group. 
\lb\ was obtained from \leda, $r_e$ in the B band from RC3 \citep{rc3}, \lk\ 
(using the luminosity distance of Table \ref{table_obs}) and $r_e$ in the K band from
\twomass.
}
\end{deluxetable*}
\def\arraystretch{1.0}


\section{X-ray images} \label{sect_images}

The X-ray image of each group was examined to identify any significant
surface brightness disturbances indicating departures from hydrostatic
equilibrium. Low level X-ray disturbances like the weak signs of AGN
activity in the center of Abell 262
\citep{blanton04}, or the presence of a submerging group in the south
of RGH 80 \citep{mahdavi05}, do not seriously impact X-ray mass
determinations, provided care is taken to avoid highly
disturbed emission \citep{buot95,schindler96}. 
The images for 15 objects in the sample
are shown in Fig.\ref{fig.1}: for objects which have \xmm\ data we
show the combined MOS1 and MOS2 image in the 0.5-2.0 keV band. For
those objects with only \chandra\ data (3 out of 16) we did the
following: For RXJ 1159.8+5531 and MS 0116.3-0115 we display the
0.5-10 keV ACIS-S3 image, while for ESO 3060170 we show the 0.5-10 keV
ACIS-I image. The images were processed to remove point sources using
the \ciao\ tool {\em dmfilth}, which replaces photons in the
vicinity of each point source with a locally estimated background. The
images where then flat-fielded with a 1.7 keV exposure map for
\chandra\ images and a 1.25 keV exposure map for \xmm.  Then we
smoothed the images with a 5\arcsec\ Gaussian for \chandra\ and
a 16\arcsec\ Gaussian for \xmm\ to make large-scale structure more
apparent. For both the \chandra\ and \xmm\ images of NGC 5044 we refer
the reader to
\citet{buot03a}.

None of the objects show obvious large scale disturbance in their
X-ray emission. The only notable substructure is the infalling
subgroup in the southern region of RGH 80, evident as a tail of
enhanced emission. Some disturbance is also present in the core of
RGH~80 as revealed by our \chandra\ image.  We masked in our analysis
the region of enhanced X-ray emission associated with the subgroup.
For other systems some low-amplitude, small-scale disturbances are
present, such as a surface brightness discontinuity, reminiscent of a
cold front, in the NW of IC 1860; the cavities in the central 10
kpc of Abell 262, as revealed by the \chandra\ image presented in
\citet{blanton04}; and the filamentary structures and possible cavities
in NGC 5044 \citep{buot03a}.  The ``cooling wake'' discussed in the
\xmm\ image of NGC 5044 by \citet{fino06}, is simply the bright SE arm of the finger-like
structure caused by the cavities.  Hint of cavities have been detected in NGC 4325 
\citep{russel07} and there are signs of sloshing in the core of MKW 4.
We asses the impact of these
features in \S~\ref{sect_systematics_asym}.

\begin{figure}[th]
\begin{center} \includegraphics[height=0.33\textheight,width=0.4\textwidth,angle=-90]{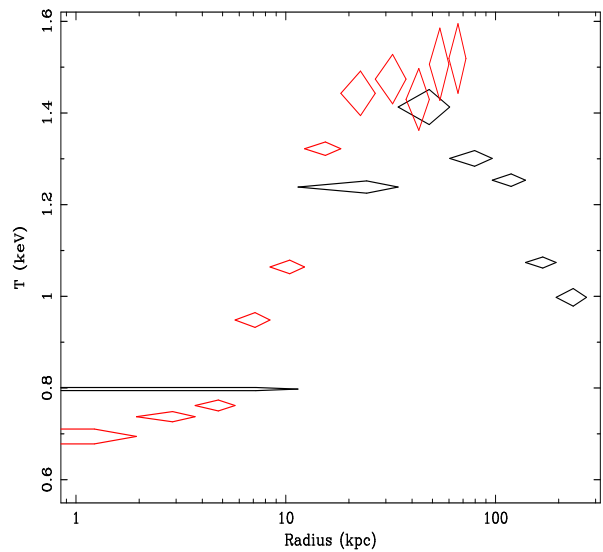} \caption{Temperature profile for the NGC 533 group derived from \xmm\ data (black) and from \chandra\ data (red). The inner two \xmm\ bins have not been considered in the derivation of the mass profile.}  \label{fig.2}
\end{center}
\end{figure}

\section{Spectral analysis} \label{sect_spectra}

We extracted spectra in concentric circular annuli located at the
X-ray centroid computed within a radius of 30\arcsec, with the initial
center on the peak of the X-ray emission. The widths of the annuli
were chosen to have approximately equal background-subtracted counts
and to have a minimum width of 60\arcsec\ for \xmm\ to avoid
under-sampling of the PSF. For \chandra, given the better PSF, the
widths of the annuli, in practice, were only limited by count
statistics.  The spectra were re-binned to ensure a signal-to-noise
ratio of at least 3 and a minimum 20 counts per bin (necessary for the
validity of the $\chi^{2}$ minimization method).  We fitted the
background-subtracted spectrum with an \apec\ thermal plasma modified
by Galactic absorption \citep{dick90} to each annulus.  The free
parameters are temperature, normalization (proportional to emission
measure), and the abundances Fe, O and, when possible, Si and S. The
impact of unresolved point sources, in particular LMXB in the central
galaxy, was taken into account by adding a 7.3 keV bremsstrahlung
component for all annuli within the twenty-fifth magnitude isophote
(\dtwentyfive) of the central galaxy, taken from \leda. (This model
gives a good fit to the composite spectrum of the detected sources in
nearby galaxies, \citealt{irwin03}.)  This is particularly relevant
for the inner \xmm\ annuli, where in general point sources are not
detected.  The spectral fitting was performed with \xspec\ \citep[ver
11.3.1,][]{xspec}. We estimated the statistical errors on the fitted
parameters by simulating spectra for each annulus using the best
fitting models and then fitted the simulated spectra in exactly the
same manner as done for the actual data. From 20 Monte Carlo
simulations we compute the standard deviation for each free parameter,
which we quote as the ``1 $\sigma$'' error \citep[these error
estimates generally agree very well with those obtained using the
standard $\Delta\chi^{2}$ approach in \xspec, e.g.,][]{humphrey05a}.

If an object has been observed by both \chandra\ and \xmm\ we selected
for our final analysis only the \chandra\ data in the inner core
region were the temperature rises outward from the center.  The \xmm\
spectra extracted in wide annuli are not as well fitted by a single
temperature emission model as are the \chandra\ spectra in narrower
annuli, suggesting that departures from single temperature emission in
the projected spectra stem primarily from the steep radial temperature
gradient present in the core, as shown in Fig.\ref{fig.2} for NGC 533.
The better \chandra\ PSF also allows us to exclude point sources
undetected with \xmm, in particular LMXB in the central
galaxy. Unresolved LMXB still affect the
\chandra\ data, but to a much lesser extent than \xmm\ data. This component is 
evident as an excess at energies greater than $\sim3$ keV and can
contribute, if neglected, to the multiphase appearance of \xmm\
spectra.

\section{Mass modeling}\label{sect_mass}

To calculate the gravitating mass distribution we solve the equation
of hydrostatic equilibrium assuming spherical symmetry. By requiring
spherical symmetry we obtain spherically averaged mass profiles which
allows us to test the spherically averaged DM profiles obtained from
cosmological simulations and to facilitate comparison to previous
observational studies.

Following the approach adopted in \citet{humphrey06a}, we assume
parametrizations for the temperature and mass profiles to calculate
the gas density assuming hydrostatic equilibrium,
\begin{equation}
\rho_g(r)=
\rho_{g0}\frac{T_0}{T(r)}\exp\left(\frac{-\mu
m_A G}{k_B}\int_{r_0}^{r}\frac{M\, dr}{r^2T}\right), \label{eqn_hydrostatic_rho}
\end{equation}
where r is the radius from the center of the gravitational potential,
\rhog\ is the gas density, $\rho_{g0}$ and $T_0$ are density and temperature
at some ``reference'' radius $r_0$, $k_B$ is Boltzmann's constant, G
is the universal gravitational constant, $m_A$ is the atomic mass unit
and $\mu$ (taken to be 0.62) is the mean atomic weight of the gas. The
$\rho_g(r)$ and $T(r)$ profiles are fitted simultaneously to the
data to constrain the parameters of the temperature and mass
models. Since the gravitating mass also contains the gas mass,
eq. (\ref{eqn_hydrostatic_rho}) is solved iteratively for \rhog.

This ``parametric mass method'' is the principal approach employed in
this study. We assess systematic errors in this adopted method in \S
\ref{sect_systematics_method} by comparing to results obtained using other
solutions to the hydrostatic equilibrium equation. Firstly, rather than
solving for the gas density, we can solve for the temperature,
\begin{equation}
T(r)=
T_0\frac{\rho_{g0}}{\rho_g(r)} - \frac{\mu m_A G}{k_B\rho_g(r)}
\int_{r_0}^{r}\frac{\rho_g M\, dr}{r^2},
\label{eqn_hydrostatic_t}
\end{equation}
which provides an alternative implementation of the ``parametric mass
method''. Note that in both cases the parameters of the mass model are
obtained from fitting the gas density and temperature data. The
goodness-of-fit for any mass model (e.g., NFW) can be assessed
directly from the residuals of the fit.  Secondly, we use the more
traditional formulation of the hydrostatic equilibrium equation
\citep{mathews78},
\begin{equation}
\label{eqn_hydrostatic_classic}
M(<r) = r{k_BT(r) \over G\mu m_A} \left( -{d\ln \rho_g \over d\ln
r} - {d\ln T \over d\ln r} \right)
\end{equation}
which involves parametrizing independently $\rho_g$ and $T$ using simple
functional forms in order to evaluate the derivatives in
Eq. (\ref{eqn_hydrostatic_classic}). Since, however, the mass profile
itself is not parametrized, we denote this traditional approach
a ``non-parametric mass method''. Since the mass profile itself is
produced by this method, if one wants to evaluate the success of a
particular mass model (e.g., NFW) then additional fitting is
required. Consequently, following previous studies
\citep[e.g.,][]{lewi03a} for each annulus we assign a three dimensional radius 
value $r \equiv [(R_{\rm out}^{3/2} + R_{\rm in}^{3/2})/2]^{2/3}$,
where $R_{\rm in}$ and $R_{\rm out}$ are respectively the inner and
outer radii of the (two-dimensional) annulus. At each radius $r$ we
calculate the total enclosed gravitating mass $M(<r)$ according to the
equation of hydrostatic equilibrium.  The errors on the resulting mass
``data points'' were estimated from the Monte Carlo simulations used
to estimate the errors for density and temperatures (\S
\ref{sect_spectra}), giving a set of mass values at each radius. From
those we calculate the standard deviation which we quote as the
``1$\sigma$'' error for this method.  To analyze the shape of the mass
profiles, we fitted parametrized models to the mass values.

There are several reasons why we adopt equation
[\ref{eqn_hydrostatic_rho}] instead of equation
[\ref{eqn_hydrostatic_classic}] for our analysis. Firstly, as noted
above, the ``parametric mass method'' allows a particular mass model
to be constrained immediately by the gas density and temperature data
and the goodness-of-fit of the mass model can be assessed in a
straightforward manner. Secondly, despite the high-quality X-ray data
provided by \chandra\ and \xmm, it is still not possible to compute
accurate derivatives of the temperature and density profiles at each
radius. Consequently, smooth models must be fitted to the entire
radial profiles, which may not produce physical solutions to equation
[\ref{eqn_hydrostatic_classic}]; e.g., jagged, non-monotonically
increasing mass profiles. Thirdly, we analyze the projected
temperatures and densities which requires the models for the gas
density (density squared, see below) and temperature to be projected
along the line-of-sight. This requires evaluating the models at least
out to the virial radius, well outside the outer radius of most of the
groups in our sample. By using the ``parametric mass method'' any
extrapolation of the gas density (equation
[\ref{eqn_hydrostatic_rho}]) or temperature (equation
[\ref{eqn_hydrostatic_t}]) is performed consistently within the
context of the assumed mass profile. It is for this last reason we
have a slight preference for using equation
[\ref{eqn_hydrostatic_rho}] over equation [\ref{eqn_hydrostatic_t}]
for this study.  Nevertheless, despite these differences, we find that
the different approaches to the mass modeling represented by the three
equations give consistent results, within the errors, for the global
halo properties (see also \S \ref{sect_systematics_method}).

For our default analysis we projected parametrized models of the
three-dimensional quantities, $\rho_g^2$ and $T$, and fitted these
projected models to the results obtained from our analysis of the data
projected on the sky (see \S \ref{sect_spectra}). The models have been
integrated over each radial bin (rather than only evaluating at a
single point within the bin) to provide a consistent comparison. They
also have been projected along the line of sight including the radial
variation in the plasma emissivity \emiss, using a model fitted to the
observed \zfe\ profile. We provide a review of the projection of
spherical coronal gas models for comparison to X-ray spectral data in
Appendix \ref{sect_deproj_equations}.

On the left panels of Fig.\ref{fig.3}, Fig.\ref{fig.4},
Fig.\ref{fig.5} and Fig.\ref{fig.6} we show the radial profiles of
the emission-weighted projection of \rhog$^{2}$ (i.e., proportional to
the $norm$ parameter of the \apec\ model divided by the area of the
annulus) along with the best-fitting model and residuals; in the
central panels we show the radial profiles of the measured $T$ along
with the best fitting emission-weighted projected model and residuals.

\subsection{Gas density models}

We considered two models for fitting of the gas density
profile: the $\beta$ model \citep{beta} and a cusped $\beta$ model
\citep{pratt02,lewi03a}, the latter of which is a modified $\beta$
model allowing for steepening of the profile in the inner regions ($r
<r_c$). This model was introduced to account for the sharply peaked
surface brightness in the centers of relaxed X-ray systems. This model
has now been widely used for both low-redshift \citep[e.g,][]{pratt02}
and high-redshift \citep[e.g.,][]{kotov05} clusters. It is preferred
with respect to the double-$\beta$ model \citep[e.g.,][]{mohr99}
because the two models give fits of comparable quality, while the
cusped $\beta$ model has one less free parameter. The cusped $\beta$
model is also better behaved in the mass determination using the
``non-parametric method'' defined above (eqn.\
\ref{eqn_hydrostatic_classic}).

\subsection{Temperature models}

The projected temperature profiles for our groups show a large degree
of similarity.  We adopted several parametrizations that have enough
flexibility for each system to describe the temperature profile
reasonably well and to explore the sensitivity of our results to the
particular functional form.  The analytic models we construct are the
following: 

\begin{itemize}

\item Smoothly joined power laws:


\begin{displaymath}
T(r) =  {1 \over \left[ \left( \frac{1}{t_1(r)} \right) ^{s} +
\left( \frac{1}{t_2(r)} \right) ^{s}\right]^{\frac{1}{s}}}
\end{displaymath}
\begin{equation} 
t_i(r) =  T_{i,100} \left( {r \over 100 \rm{kpc}} \right)^{p_i} i=1,2
\end{equation}

\item Power laws mediated by an exponential:


\begin{displaymath}
T(r)  =  T_0 + t_1(r)e^{-(\frac{r}{r_p})^{\gamma}}+ t_2(r)\left(1-e^{-(\frac{r}{r_p})^{\gamma}} \right) 
\end{displaymath}
\begin{equation} 
t_i(r)  =  T_{i} \left( {r \over r_0} \right)^{p_i}  i=1,2
\end{equation}

\item The \citet{alle01c} rising profile joined to a 
falling temperature profile by an exponential cut-off,


\begin{displaymath}
T(r)  =  t_1(r)e^{-(\frac{r}{r_p})^{\gamma}}+ t_2(r)\left(1-e^{-(\frac{r}{r_p})^{\gamma}} \right) 
\end{displaymath}
\begin{displaymath}
t_1(r)  =  a + T_1\left[ \frac{\left({r \over r_1}\right)^{p_1}}{1+\left({r \over r_1}\right)^{p_1}} \right]
\end{displaymath}
\begin{equation} 
t_2(r)  =  b + T_2\left[ \frac{1}{1 +\left({r \over r_2}\right)^{p_2}} \right].
\end{equation}

\end{itemize}

The third (``RiseFall'') model has been adopted in particular for
temperature profiles showing an inner core flattening like NGC 533,
NGC 4325 and NGC 5044, while the first two models provide comparable
fits to the general profile. We will assess how different choices of
temperature profile, together with density profiles, affect our mass
measurements in \S \ref{sect_systematics_method}.

\subsection{Mass models}\label{sect_mass_models}

We compute the total gravitating mass as the sum of DM, stars, and hot
gas: $M_{DM} + M_{stars} + M_{gas}$. For this study we only consider
the contribution of the central galaxy to the stellar mass.  The X-ray
data provide a direct measurement of the hot gas density and therefore
of $M_{gas}$.  We tested the following mass models against the data:

\begin{itemize}
\item $M_{DM}$= NFW, $M_{stars}$= 0: A single NFW model to
investigate scenarios like the ones of \citet{loeb03} and
\citet{elzant04} and the effect of the omission of the stellar mass,
if present, on the derived concentration parameter.

\item $M_{DM}$= NFW, $M_{stars}$= deV. A NFW model plus a model for a stellar 
component. We adopted a de Vaucolueurs stellar mass potential using
the ana\-ly\-tical approximation to the deprojected \sersic\ model of
\citet{prugniel97} with $n=4$. The de Vaucolueurs profile is a good
description of the stellar light distribution even for objects which
follow the more general Sersic profile with Sersic index $n \neq 4$
\citep[see appendix A of][]{padma04}. The de Vaucolueurs effective radius $r_e$ is 
measured in the K-band by the Two Micron All 
Sky Survey (\twomass\ ) as listed in the Extended Source Catalog 
\citep[][see Tab.\ref{tab:optical}]{jarrett00}. 
We refer to this model as NFW+stars.

\item $M_{DM}$= NFW*AC, $M_{stars}$= deV. A NFW component modified by the adiabatic 
contraction model of \citet{gnedin04}\footnote{The adiabatic
contraction code we used was made publicly available by Oleg Gnedin
at: http://www.astronomy.ohio-state.edu/$\sim$ognedin/contra/} plus a
de Vaucolueurs component for the stellar mass, to explore the
importance of baryon condensation in the central galaxy for the DM
halo profile. We refer to this model as NFW*AC+stars.

\item Finally we also examined the recently suggested Sersic-like profile 
\citep[][hereafter N04]{nav04} which should be a better
parametrization of the innermost regions of CDM halos.
\end{itemize}

To obtain the true virial radius and virial mass (and concentration)
we initially take $r_{\Delta}$ and $M_{\Delta}$ obtained for the DM
component, where $\Delta$ corresponds to the over-density level (2500,
1250, 500) closest to the radial range covered by the data and listed
in Table \ref{tab:virial_pot_gas}. Then we added $M_{gas}$ and
$M_{stars}$ to $M_{DM}$ to give a new $M_{\Delta}$.  A new
$r_{\Delta}$ is then computed, and the process repeated, until
$r_{\Delta}$ changes by $<0.001\%$. (We note that in our previous
studies by \citealt{humphrey06a} and \citealt{zappacosta06} we also
computed the virial radius appropriate for all of the mass
components.) The values thus obtained have then been converted to
various over-densities (in particular the virial over-density,
listed for each object in Table \ref{table_obs}) in Table
\ref{tab:delta} by using the formula provided by \citet{hu03}
appropriate for an NFW halo.  We prefer this procedure for
extrapolating the mass and concentration to $\Delta\approx 101$ (for
comparison with theoretical models) because it does not involve also
extrapolating $M_{gas}$. We find that the extrapolated values for the
gas mass are sensitive to the radial range over which the density
profile is fitted (see \S
\ref{sect_systematics_range}).

\begin{figure*}[t]
\centerline{\includegraphics[scale=.8]{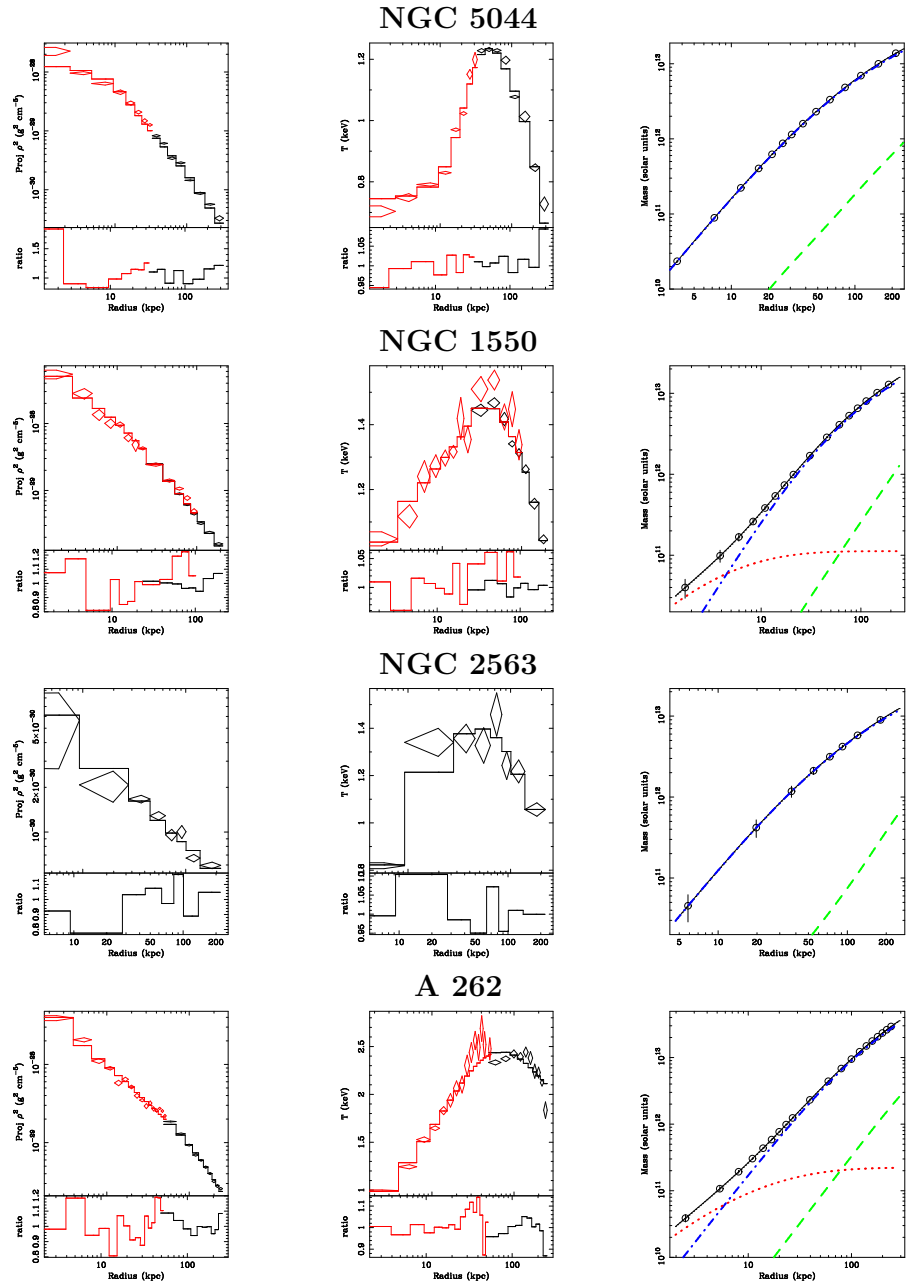}}
\vspace*{-55mm}
\caption{\label{fig.3} Results for the emission-weighted projection of
the gas density squared (left panels), the emission-weighted projected
temperature (central panel), and the total gravitating mass (right
panel) for NGC 5044, NGC 1550, NGC 2563 and A262. In the temperature
and density plots, red symbols corresponds to \chandra\ data, while
black symbols corresponds to \xmm\ data. Residuals from the best-fit
``parametric mass method'' models (\S \ref{sect_mass}) for NFW(+stars)
are also shown. In the gravitating mass plot the different mass
components are shown: DM with the dotted (blue) line, gas mass with
the dashed (green) line and stellar mass with the dotted (red) line.
Representive ``data points'' are plotted in the gravitating mass profile to
show the size of the error bars on the total gravitating mass.}
\end{figure*}


\begin{figure*}[t]
\centerline{\includegraphics[scale=.8]{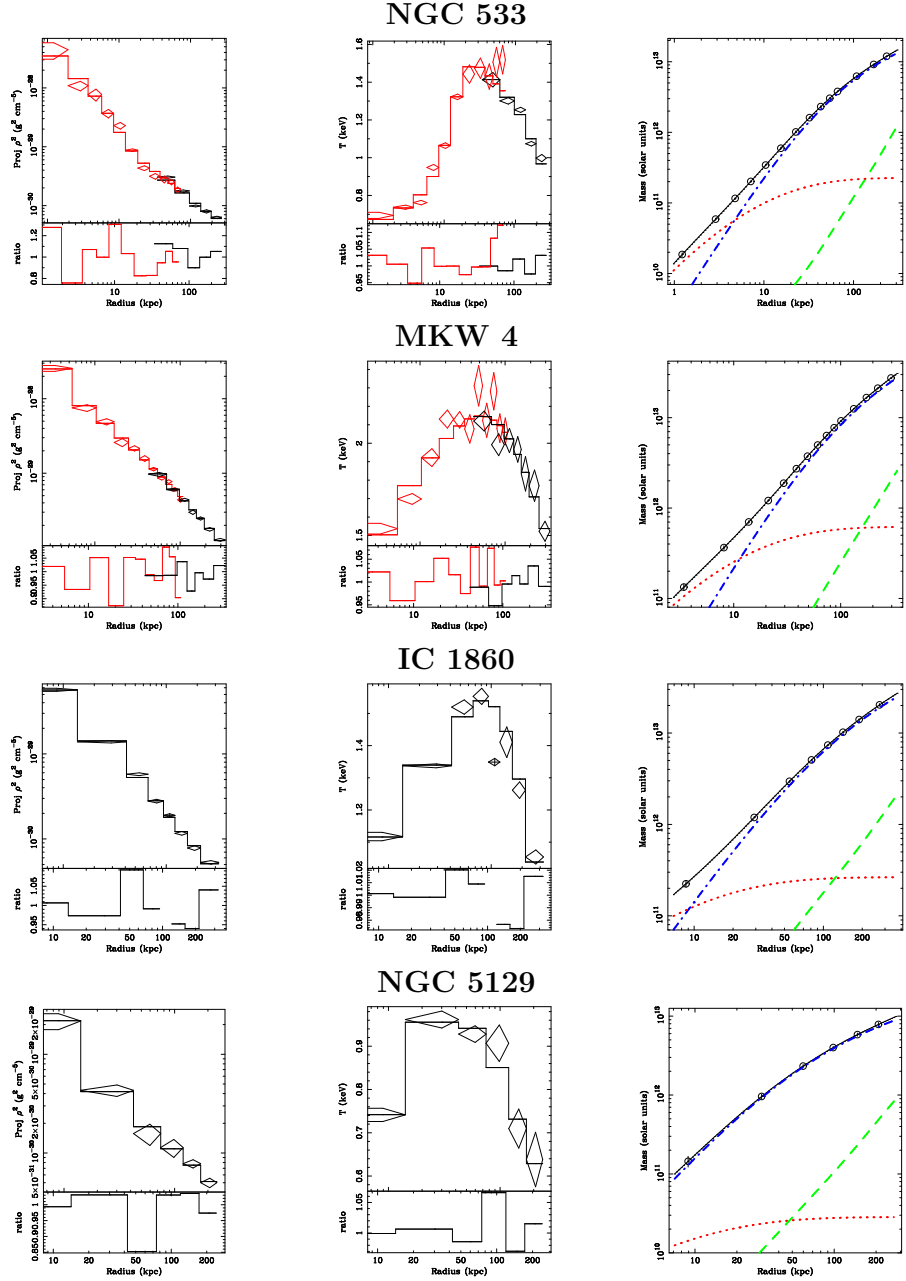}}
\vspace*{-30mm}
\caption{\label{fig.4} Same as in Fig.\ref{fig.3} for NGC 533, MKW 4, IC 1860 and NGC 5129. The crossed values for the annular bin around 100 kpc for IC 1860 indicates that the data point has not been considered in the fit. 
}
\end{figure*}
\clearpage


\begin{figure*}[t]
\centerline{\includegraphics[scale=.8]{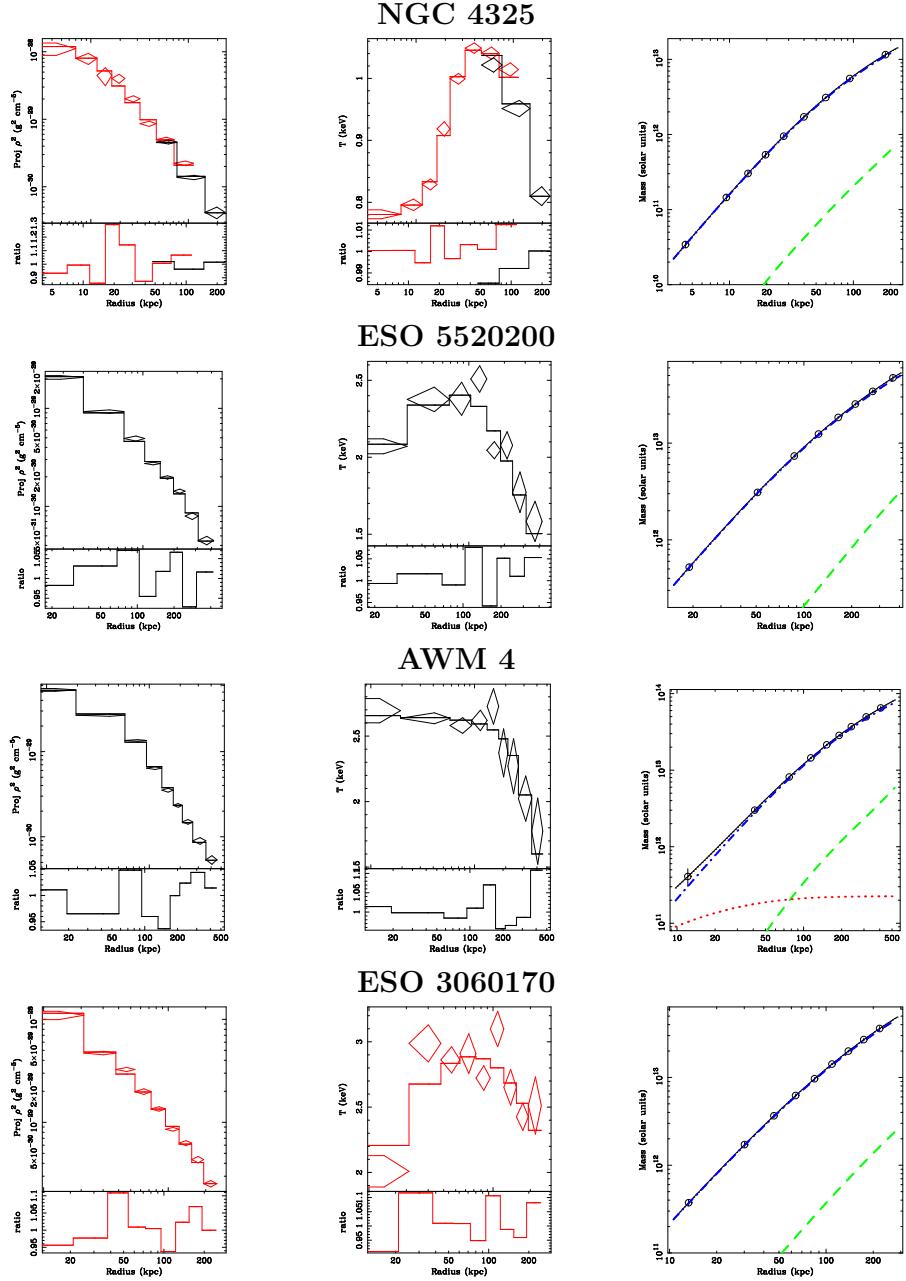}}
\vspace*{-30mm}
\caption{\label{fig.5} Same as in Fig.\ref{fig.3} for NGC 4325, ESO 5520200, AWM 4 and ESO 3060170.}
\end{figure*}


\begin{figure*}[t]
\centerline{\includegraphics[scale=.8]{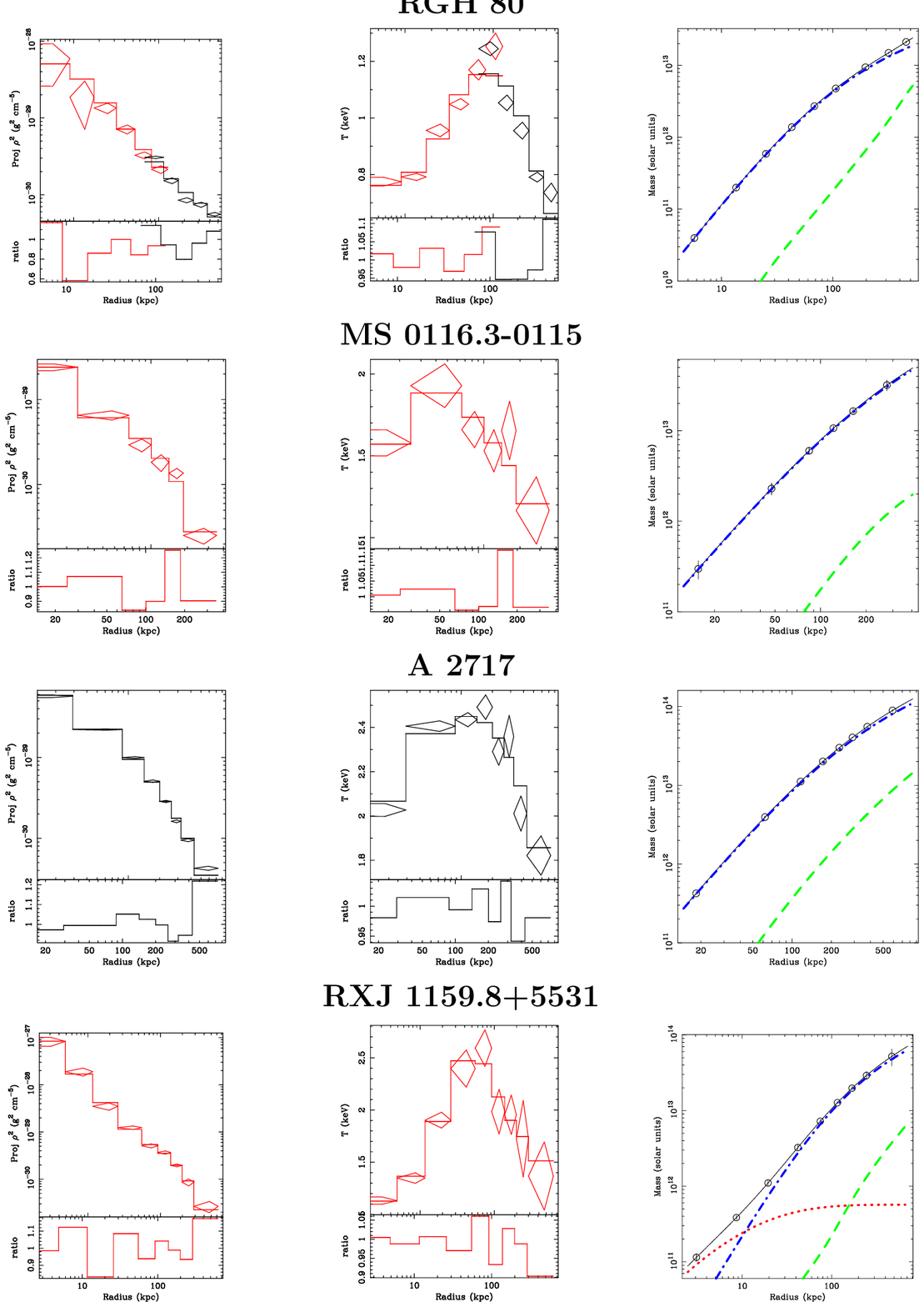}}
\vspace*{-30mm}
\caption{\label{fig.6} Same as in Fig.\ref{fig.3} for RGH 80, MS 0116.3-0115, A 2717 and RXJ 1159.8+5531.}
\end{figure*}

\section{Results} \label{sect_results}

\subsection{Mass-fitting results}\label{fit_goodness}

We tested the different mass models listed in
\S~\ref{sect_mass_models} against the data. In the following analysis
we obtain the best fit by minimizing $\chi^2$. Although our best fit
models are not formally acceptable, the major contributions to
$\chi^2$ stem generally from the inner data points ($\la 10$ kpc),
where the errors in both temperature and density are small. It is not
expected that the DM halo of each system will be perfectly fitted by
an NFW profile \citep[e.g.,][]{tasitsiomi04}. Consequently, we also
quote the values of the maximum fractional deviation $\rm{dvi_{max}}$,
which gives equal weight to all radial bins, as a figure of goodness
of fit, in addition to $\chi^2$ in Table \ref{table_chisq}.  The
quantity $\rm{dvi_{max}}$ is routinely used in the fits to halos
formed in numerical simulations; e.g., \citet{jing00} proposes that
$\rm{dvi_{max}} < 0.30$ represents a good fit of the NFW model.  The
results for the best fit NFW or NFW+stars model are listed in Table
\ref{tab:virial_pot_gas} at the appropriate over-density co\-ve\-red by the
data. 

Our basic result is that the NFW model, sometimes benefiting from an
additional component from the stellar mass in the central galaxy, is a
good overall description of the mass profiles. While the formal
quality of the fits, as noted above, are generally not acceptable, the
fractional deviations of the fits are typically $\sim 10\%$. The
largest maximum deviation is observed for NGC 5044 within its central
radial bin ($<3$~kpc), where the \chandra\ ima\-ge shows
irregularities presumably associated with AGN feedback
\citep{buot03a}. At all other radii the de\-via\-tions are $\la 10\%$ for
NGC 5044.

The stellar mass component is not uniformly required. When using the
NFW+stars model, only 8 objects of the 16 in the sample show an
improvement in the fit. The improvement of the fit is judged by
considering a reduction in $\chi^{2}$ and a reduction in fractional
residuals.  This provides a quantitative assessment of the improvement
of the fit even if the final $\chi^{2}$ is still not formally
acceptable. For example an NFW fit to the MKW4 mass profile gives
$\chi^{2}$/dof=58/25 with a $\rm{dvi_{max}}$ of 0.60 arising from the
central density bin, while the best fit NFW+stars gives
$\chi^{2}$/dof=34/24 with a $\rm{dvi_{max}}$ of 0.13, because the
inner data points are better modeled.

Moreover the amount of improvement is sensitive both to the number of
data points sampling the inner $\approx 20$ kpc (where the stellar
mass is expected to make a substantial contribution to the total mass
budget) and to the luminosity of the central galaxy.  For this purpose
it is instructive to examine those systems that require stellar mass
and have both \chandra\ and \xmm\ data --- NGC 1550, A 262, NGC 533
and MKW 4. By fitting only the \xmm\ data, with its coarser binning at
small radius, we can assess the importance of having high-resolution
\chandra\ data for detecting a stellar mass component. When fitting
only the \xmm\ data for these systems the evidence for a stellar mass
component is weaker, and the inferred amount of stellar mass less,
than for the simultaneous \chandra\--\xmm\ fits. The amount of stellar
mass inferred is always larger when the \chandra\ data are included.
In 3 of the 4 cases, the derived concentration value does not change
within the $1-2\sigma$ errors. The exception is A262 for which
$c_{\Delta}$=$1.2\pm0.1$ is obtained using only \xmm\ and
$c_{\Delta}$=$2.1\pm0.2$ is found for the simultaneous
\chandra\--\xmm\ fits.

It follows that for systems having only \xmm\ data it is necessary to
obtain high-quality \chandra\ observations to make a reliable
detection of stellar mass.  There is clear
failure to detect stellar mass in 3 objects in our sample that
are adequately sampled by \chandra\ observations: NGC 5044, NGC 4325
and RGH 80.

The omission of the stellar component in the mass fits leads to biased high
concentrations \citep{mamon05a,humphrey06a}, but the relevance of the
bias depends on the number of data points sampling both the stellar
component (dominant in the inner $\sim$ 20 kpc) and DM component.  The
objects in our sample have adequate sampling of the DM component at
relatively large radii, but the stellar component is well sampled
($\sim$3 data bins in the inner 20 kpc) only when \chandra\ data are
present. As a consequence, the bias is more pronounced when \chandra\
data are included.  This effect is most evident for MKW4.  An NFW fit
to the \xmm\ data for MKW4 gives $c_{\Delta}=5.8\pm0.3$ while an
NFW+stars fit gives $c_{\Delta}=4.8\pm0.4$. If we use
\chandra\ and \xmm\ data, then the fit is driven by the increased number of
data points within 20 kpc. Fitting an NFW model yields,
$c_{\Delta}=6.8\pm0.2$, which represents a 58\% increase over our best
fit NFW+stars value, $c_{\Delta}=4.3\pm0.3$ (see Table
\ref{tab:virial_pot_gas}).  For the remaining objects, fitting only
the NFW model, when NFW+stars is required, returns a $c_{\Delta}$
biased high in the range 38\% (A262) to 10\% (NGC 533).  As expected,
the bias is generally less for our groups-clusters ($M >
10^{13}$~\msun) than obtained for the elliptical galaxies-groups
($M\la 10^{13}$~\msun) analyzed by \citet{humphrey06a}.

In order to explore the presence and relevance of adiabatic
contraction we fitted an NFW*AC+stars model to the 8 objects which
require a stellar mass component, because only for these objects is
the AC model potentially relevant. The quality of the fits is not
improved by the introduction of AC (see Table
\ref{table_chisq}). Because the AC model increase the cuspiness of the
DM profile, we find that the stellar mass (and the derived stellar
mass-to-light ratios, see Table \ref{table_mass_to_light}) were
considerably lower for the NFW*AC+stars model than for the NFW+stars
model. Because less stellar mass is obtained for the AC models, the
derived $c_{\Delta}$ values increase by 10-40\% compared to NFW+stars. Two
exceptions are MKW 4 and RXJ 1159.8+5531, for which AC increases
$c_{\Delta}$ to $7.1\pm0.4$ and $9.6\pm1.9$ respectively.  The quality
of the NFW*AC+stars fits is considerably worse in these two cases
compared to NFW+stars.  The $M_{\Delta}$ values obtained for the
NFW*AC+stars model are also lower by 5-20\%, with a maximum of 33\%
for RXJ 1159.8+553.

Finally, we examined the N04 model. We explored N04+stars because in
our previous analysis of the cluster Abell 2589
\citep{zappacosta06} N04 allowed for an increased contribution from
stellar central mass components (with values of the Sersic parameter
$\alpha \sim 0.4$).  Even if we left the Sersic parameter $\alpha$
free, the fit improved only in few cases -- and only in two, A 2717
($\chi^2$/dof=7/4) and IC 1860 ($\chi^2$/dof=7/4) was the improvement
superior to 90\% according to the F-test. The inferred values of
$\alpha$ for the sample were quite large and incompatible with the
mean value of $0.17\pm0.03$ for CMD halos
\citep{nav04}. If we fixed the value of $\alpha$ at 0.17 
the fits did not improve.

\begin{deluxetable*}{lccc}
\tablecaption{Quality of the mass fits\label{table_chisq}}
\tablecolumns{6}
\tablehead{ \colhead{Group} & \multicolumn{2}{c}{$\chi^2$ } &
\colhead{$\rm{dvi_{max}}$  } \\
& \colhead{NFW(+stars)} & \colhead{NFW*AC+stars} &\colhead{NFW(+stars)} 
}
\startdata
NGC 5044        & 228/20 & \dotfill   & 0.83   \\ 
NGC 1550        & 66/33  & 74/33      & 0.22   \\
NGC 2563        & 18/7   & \dotfill   & 0.23   \\
A 262           & 116/37 & 134/37     & 0.19   \\
NGC 533         & 81/20  & 85/20      & 0.30   \\
MKW 4           & 34/24  & 63/24      & 0.13    \\
IC 1860         & 14/5   & 25/5       & 0.11    \\
NGC 5129        & 3/2    & 3/2        & 0.15    \\
NGC 4325        & 12/9   & \dotfill   & 0.29    \\
ESO 5520200     & 18/8   & \dotfill   & 0.08   \\
AWM 4           & 13/9   &  14/9      & 0.11   \\
ESO 3060170     & 25/9   & \dotfill   & 0.12   \\
RGH 80          & 71/9   & \dotfill   & 0.42   \\
MS 0116.3-0115  & 6/5    & \dotfill   & 0.16   \\
A 2717          & 32/9   & \dotfill   & 0.22   \\
RXJ 1159.8+5531 & 5/7    &  13/7      & 0.17   \\
\enddata
\tablecomments{Various indicators of quality of fit for the mass models
discussed in the text.\\
$\chi^{2}$ refers to the $\chi^{2}$/dof of the fits to the density and 
temperature profiles used to infer the parameters of the mass model in the gas 
potential approach.\\
$\rm{dvi_{max}}$ refers to the maximum fractional deviation on the fits to the 
density and temperature profiles used to infer the parameters of the mass 
model in the gas potential approach.\\
Values of $\chi^2$ for the NFW*AC+stars model are reported only for the objects showing an excess over the NFW fit due to stellar mass.
}
\end{deluxetable*}



\def\arraystretch{1.25}
\begin{deluxetable*}{llllcccccccc}
\setlength{\tabcolsep}{0.05in}
\tablecaption{Results for the NFW virial quantities at selected overdensity \label{tab:virial_pot_gas}}
\tabletypesize{\scriptsize}
\tablehead{
\colhead{Group\hspace*{15mm}} &
\colhead{$\Delta$} &
\colhead{$r_{s}$} &
\colhead{$c_{\Delta}$} &
\colhead{$r_{\Delta}$} &
\colhead{$M_{\Delta}$} &
\colhead{$M_{gas,\Delta}$} &
\colhead{$f_{gas,\Delta}$} &
\colhead{$M_{DM,\Delta}$} &
\colhead{$M_{\star,\Delta}$} \\
 & & \colhead{(kpc)} & & \colhead{(kpc)} &\colhead{($10^{13}\,M_\odot$)} & \colhead{($10^{12}\,M_\odot$)} &  &\colhead{($10^{13}\,M_\odot$)} & \colhead{($10^{10}\,M_\odot$)} & }
\startdata
NGC 5044  & 1250 & $77\pm2$  & $ 3.8\pm0.1$ & $295\pm2$  & $1.85\pm0.04$  & $1.21\pm0.02$ & $0.065\pm0.001$  & $1.72\pm0.04$  &       &   \\ 
NGC 1550  & 2500 & $48\pm4$  & $ 4.5\pm0.3$ & $215\pm2$  & $1.42\pm0.03$  & $1.02\pm0.02$ & $0.072\pm0.001$  & $1.31\pm0.03$  & $11.2\pm4.1$  &   \\
NGC 2563  & 2500 & $76\pm22$ & $ 2.4\pm1.0$ & $185\pm5$  & $0.92\pm0.08$  & $0.31\pm0.03$ & $0.034\pm0.001$  & $0.89\pm0.08$  &       &   \\
Abell 262  & 2500 & $141\pm16$ & $2.1\pm0.2$ & $292\pm4$  & $3.59\pm0.14$  & $2.60\pm0.08$ & $0.072\pm0.001$  & $3.31\pm0.13$  & $22.1\pm4.5$  &   \\
NGC 533   & 1250 & $43\pm4$  & $ 6.1\pm0.5$ & $262\pm2$  & $1.30\pm0.04$  & $0.87\pm0.02$ & $0.067\pm0.001$  & $1.19\pm0.04$  & $22.4\pm2.2$  &   \\
MKW4      & 1250 & $81\pm7$  & $ 4.3\pm0.3$ & $353\pm4$  & $3.21\pm0.10$  & $2.84\pm0.06$ & $0.088\pm0.002$  & $2.87\pm0.10$  & $61.8\pm7.2$   &  \\
IC 1860   & 1250 & $101\pm12$ & $ 3.2\pm0.3$ & $319\pm6$  & $2.36\pm0.13$  & $1.56\pm0.05$ & $0.066\pm0.002$  & $2.18\pm0.12$  & $26.4\pm6.3$  &   \\ 
NGC 5129  & 1250 & $43\pm10$ & $ 5.2\pm0.9$ & $226\pm7$  & $0.84\pm0.07$  & $0.58\pm0.06$ & $0.069\pm0.003$  & $0.78\pm0.07$  & $2.8^{+6.7}_{-2.8}$  &   \\
NGC 4325  & 2500 & $75\pm18$ & $ 2.8\pm0.4$ & $208\pm8$  & $1.32\pm0.16$  & $0.66\pm0.03$ & $0.050\pm0.004$  & $1.26\pm0.16$  &       &   \\
ESO5526020  & 1250 & $171\pm27$ & $2.5\pm0.3$ & $422\pm13$ & $5.51\pm0.51$ & $3.35\pm0.18$ & $0.061\pm0.002$  & $5.17\pm0.50$   &    &    \\
AWM 4     & 1250 & $154\pm17$ & $ 3.0\pm0.3$ & $465\pm13$ & $7.38\pm0.61$ & $4.79\pm0.29$ & $0.065\pm0.003$  & $6.88\pm0.59$ & $22.5^{+24.7}_{-22.5}$ &  \\
ESO3060170  & 2500 & $162\pm54$ & $2.1\pm0.3$ & $343\pm18$ & $5.97\pm1.14$ & $3.45\pm0.17$ & $0.058\pm0.005$ & $5.62\pm1.12$   &    &    \\
RGH 80  & 500 & $78\pm8$ & $5.1\pm0.5$   & $397\pm5$ & $1.85\pm0.07$  & $2.85\pm0.11$ & $0.154\pm0.003$  & $1.56\pm0.06$  &   &   \\ 
MS 0116.3-0115  & 1250 & $202\pm115$ & $2.0\pm0.8$ & $405\pm42$ &$4.92\pm1.64$ & $1.97\pm0.19$ & $0.040\pm0.009$ & $4.73\pm1.63$ &   &   \\
Abell 2717  & 500 & $233\pm18$ & $3.0\pm0.2$ & $710\pm11$ & $10.68\pm0.51$ & $11.36\pm0.29$ & $0.106\pm0.003$  & $9.55\pm0.49$ &       &  \\
RXJ 1159.8+5531  & 500 & $104\pm77$ & $5.6\pm1.5$ & $584\pm73$ &$6.13\pm3.30$ & $5.10\pm0.41$ & $0.083\pm0.019$ & $5.57\pm3.32$ & $56.9\pm10.5$  & \\
\enddata
\tablecomments{Results for the mass profile fits. $\Delta$ refers to the overdensity chosen for the object, as the closest to the outer radius of the data. $r_{s}$
is the scale radius of the NFW profile.}
\end{deluxetable*}
\def\arraystretch{1.0}

\subsection{Stellar Mass-to-Light ratios}\label{stellar_ml}

\begin{deluxetable*}{llll}
\tablecolumns{5}
\tablecaption{Stellar mass-to-light ratios\label{table_mass_to_light}}
\tablehead{ \colhead{Galaxy} & \colhead{\lk/\lb} &
    \multicolumn{2}{c}{Fitted \mstar/\lk\ (\msun/\lsun)} \\
    \colhead{} & \colhead{} & \colhead{NFW+H90} &  \colhead{NFW*AC+H90} 
}
\startdata
NGC\thin1550     & 4.8 & $0.53\pm0.20$ & $0.24\pm0.01$ \\
NGC\thin 708     & 10.7 & $0.54\pm0.11$ & $0.14\pm0.10$ \\
NGC\thin 533     & 4.9 & $0.36\pm0.03$ & $0.11\pm0.01$ \\
NGC\thin 4073    & 4.6 & $0.86\pm0.10$ & $0.33\pm0.04$ \\
IC\thin 1860     & 7.2 & $0.60\pm0.14$ & $0.26\pm0.01$ \\
NGC\thin 5129    & 4.1 & $0.06^{+0.13}_{-0.06}$ & $0.05\pm0.01$ \\
NGC 6051         & 7.6 & $0.30^{+0.33}_{-0.30}$ & $0.18\pm0.01$ \\ 
2MASSX J11595215 & 4.4 & $0.55\pm0.10$ & $0.40\pm0.05$ \\
\enddata
\tablecomments{K-band stellar mass-to-light ratios for the central galaxy 
measured from our fits to the data using the NFW+stars and the NFW*AC+stars 
models.} 
\end{deluxetable*}

Using the stellar mass derived from our fits we calculated the stellar M/L 
ratios (\mstar/L) for the central galaxy.
The optical luminosities have been calculated in the K$s$-band, following 
\citet{kochanek01} and \citet{linmohr04}, using (1) the 20 mag $\rm{arcsec}^{-2}$
isophotal elliptical aperture magnitude, (2) the value of the Galactic
extinction provided by \ned, (3) a $k$-correction of the form $k(z) =
-6\rm{log}(1+z)$, and (4) a correction of 0.2 mag to convert between
the total and isophotal absolute magnitudes \citep[see Appendix
of][]{kochanek01}. We compare this estimate to the total extrapolated
magnitudes listed in the Extended Catalog, finding agreement to better
than 10\%\footnote{For a discussion regarding the use of the
elliptical isophotal magnitude instead of the extrapolated total
magnitude, because it is less vulnerable to stellar contamination and
surface brightness irregularities, see the FAQ sheet for the \twomass\
Extended source catalog at
http://spider.ipac.caltech.edu/staff/jarrett/2mass/XSC/\-jarrett\_XSCprimer.html}.  
For distances we adopted the luminosity distance listed in Table \ref{table_obs}. 
Magnitudes have been converted to units of B and 
K$s$ solar luminosities using $M_B\odot = 5.48$
\citep{girardi00} and $M_{Ks}\odot = 3.34$, which follows from 
adopting $(B-V)\odot = 0.64$ and $(V-Ks)\odot = 1.50$ \citep{holmberg06}.

We use the K-band to quote \mstar/L because Near-infrared (NIR)
luminosities are much more closely correlated with the total galaxy
mass than optical luminosities
\citep{gavazzi96}. Table~\ref{table_mass_to_light} shows the best
fitting results for NFW+stars and NFW*AC+stars for those objects
requiring a stellar mass component in the mass analysis in \S
\ref{fit_goodness}.

\section{Results for individual objects} \label{sect_objects}

In the right panels of Fig.\ref{fig.3}, Fig.\ref{fig.4},
Fig.\ref{fig.5} and Fig.\ref{fig.6} we show the total gravitating
mass profiles for the objects in our sample, with the different
components (DM, gas mass and stellar mass of the central galaxy) shown
in different colors and line styles.  Details for some individual
objects, and comparison with previous results in the literature, are
provided below.

{\em NGC\thin 1550.} Our density and temperature profiles agree 
with the \chandra\ analysis of \citet{sun03}. Their fit to the 
\emph{total mass} profile within 200 kpc, not surprisingly, prefers a Moore 
profile over an NFW (in particular in the inner 10 kpc) because of the
stellar mass contribution which steepens the profile of the total
gravitating matter. Our derived DM parameters for an NFW fit are not
very different from their NFW fit to the total matter: our scale
radius $r_s$=$48\pm4$ kpc and $\delta_c$ = $7.76\pm0.56 \times 10^{4}$
are similar to their best-fitting values of 41.8 kpc and
$1.10\times10^{5}$ respectively.

{\em Abell\thin262.} The total mass profile has been analyzed using
both \chandra\ and \rosat\ data at large radii by \citet{vikh06a}, who
find a concentration, $c_{500}=3.54\pm0.30$ which is consistent within
$2\sigma$ with ours ($4.5\pm0.4$, see Table \ref{tab:delta}). Other
relevant quantities are in excellent agreement: our values of
$M_{2500}$=$3.59\pm0.14\times10^{13}$ \msun,
$f_{gas,2500}$=$0.072\pm0.001$ and $r_{500}$=$624\pm15$ kpc agree well
with their values of $3.40\pm0.50\times10^{13}$ \msun, $0.067\pm0.003$
and $650\pm21$ kpc.  Using \xmm\ data \citet{piffa05} obtained the
following parameters from a single NFW model fitted to the gravitating
matter: $r_s$=$85\pm17$ kpc, $c_{200}$=$8.6\pm1.0$,
$M_{DM,2500}$=$1.97\pm0.27\times10^{13}$ \msun\ and
$M_{gas,2500}$=$1.36\pm0.20\times10^{12}$ \msun.  These values do not
agree with ours, even when we similarly fit only the NFW model (i.e.,
no separate accounting for stars or gas) to the
\xmm\ data:  $r_s$=$174\pm10$ kpc, 
$M_{DM,2500}$=$3.39\pm0.10\times10^{13}$ \msun\ and
$M_{gas,2500}$=$2.76\pm0.06\times10^{12}$ \msun. Considering the
agreement between our results and those of \citet{vikh06a}, it is
unclear why \citet{piffa05} obtain different results for this system.

{\em NGC\thin533.} Using \xmm\ data \citet{piffa05} obtain the
following results for the NFW model fitted to the gravitating mass:
$r_s$=$37\pm3$ kpc and $c_{200}$=$12.53\pm0.55$. Under the same
conditions we obtain good agreement with their results: $r_s$=$43\pm4$
kpc and $c_{200}$=$13.0\pm0.9$.

{\em MKW\thin 4.} Our derived temperature profile shows a declining behavior
like the one obtained by \citet{vikh05a}, using \chandra\ data, by
\citet{fukazawa04}, using both \chandra\ and \xmm\ data and by \citet{fino07} 
using \xmm\ data.  However, our
profile does not agree with the relatively flat profile with higher
temperature values obtained by \citet{osul03b}.  We believe the origin
of the discrepancy probably arises from their application of the
double subtraction method to subtract the background (\S
\ref{sect_bkg}), because the emission from MKW~4 fills the entire
\xmm\ field of view.  The subtraction of
a source component artificially hardens the outermost annuli. 
A similar conclusion has been reached by \citet{fino07}.
Our mass model extrapolated to $\Delta=500$ gives, $c_{500} =
6.4\pm0.5$, $r_{500} = 527\pm8$ kpc and $M_{500} = 4.27\pm0.18 \times
10^{13}$ \msun, which do not agree with the parameters found by
\citet{vikh06a}, $c_{500} = 2.54\pm0.15$, $r_{500} = 634\pm28$ kpc and
$M_{500} = 7.7\pm1.0 \times 10^{13}$
\msun\, obtained by combining \chandra\ and \rosat\ data at large
radii (out to 550 kpc).  However, when restricting the comparison to
the radial range covered by our data, our mass
($M_{2500}=2.4\pm0.1\times10^{13}$ \msun) agrees with theirs
($M_{2500}=2.8\pm0.3\times10^{13}$ \msun).
The difference in concentrations stem primarily from a difference 
between our measured scale radius, $r_{s} = 81\pm7$ kpc and
their value of 250 kpc. As we discuss in \ref{sect_systematics_range},
a measurement of the scale radius is reliable only if it is well within 
the radial range of the data. Although by using \rosat\ data 
\citet{vikh06a} have surface brightness information out to 550 kpc,
accurate spectral information is available only with \chandra\ data,
which beyond $\sim100$ kpc (outside of the ACIS-S3 chip) are only
covering a sector of the entire radial annulus and have relatively low
S/N. The similar scale radius of 76 kpc for this object derived by
\citet{rines06} using galaxy redshifts and identifying caustics in
redshift space supports our results, although the value of $r_{500}=620$
kpc agrees better with \citet{vikh06a} ($634\pm28$ kpc) than with the present 
work ($527\pm8$ kpc). Other possible reasons for the discrepancy are the 
different mass modeling procedure or the different radial range used in the 
fit, restricted to  $r> 37$ kpc in the analysis of \citet{vikh06a}.

{\em IC\thin1860.} The group exhibits a sharp decline in surface
brightness in the NW and enhanced emission in the SE. This
particularly affects the annulus between 94 and 121 kpc, which has
been excluded from the fit. We studied the effects of this asymmetry
by dividing the annuli into two sectors. We defined the SE sector as
15-195 degrees measured from the N. The corresponding NW sector is
then 195-15 degrees. We find that the gas density profile is steeper
in the NW direction, but the lower temperature in the 91-125 kpc
annulus is caused by the cooler, higher density emission in the SE.
The radial temperature profile is quite smooth over the NW sector.
The $c_{\Delta}$ and $M_{\Delta}$ values obtained for each sector when
excluding the 94-121~kpc annulus are consistent within their $\sim
2\sigma$ errors. Including this annulus has negligible impact on the
results for the NW sector. These low level disturbances did not indicate
a significant violation of hydrostatic equilibrium, as further suggested
by the agreement of the derived \cvir\ and \mvir\ with the values 
expected from $\Lambda$CDM simulation.

{\em NGC\thin4325.} We measured an NFW scale radius $r_{s} = 75\pm18$
kpc and $M_{200}=3.01\pm0.65\times10^{13}$. The results agree with the
uncertain values obtained by \citet{rines06}; i.e.,
$M_{200}=1.5\pm1.3\times10^{13}$ and $r_{s} = 82$ kpc.
 
{\em AWM\thin 4.} This object has a remarkable temperature
profile. Unlike the other groups in our sample, the core is isothermal
as found previously by \citet{osul05}. Beyond a radius of 200 kpc we
measure a declining temperature profile with the \xmm\ data.  We find
that the source emission fills the entire field of view, contrary to
the analysis in \citet{osul05}, which reported a ``soft excess''
described by a 0.6 keV bremsstrahlung component, probably the
misinterpreted source. It is difficult to classify AWM 4 as a merging
system, given its relaxed appearance both in the X-rays and in the
optical \citep{koranyi02}. Instead, the flat temperature profile
likely reflects the influence of the powerful AGN, with radio lobes
extending out from the central galaxy NGC 6051 along the minor axis of
the galaxy to 100 kpc \citep[e.g.,][]{neumann94}.

{\em ESO\thin3060170.} Our temperature profile is best fitted by a
declining profile at large radii. However, because of the relatively
large error bars, our profile is also consistent with the flat profile
obtained by \citet{sun04} between 10 and 400 kpc with \xmm\ and
\chandra.  \citet{sun04} obtain  $c_{200}\sim8.7$ and
$M_{200}\sim1.8\times10^{13}$ \msun\ which may be compared to our
extrapolated values, $c_{200}=6.7\pm0.8$ and
$M_{200}=1.54\pm0.59\times10^{14}$ \msun.

{\em RGH\thin 80.}  The \chandra\ image clearly reveals the
peak of the X-ray emission coincident with MCG +06-29-077 and a
bright tail pointing NW -- with MCG +06-29-078 at the S edge of this
feature. This geometry was only hinted at by the \xmm\ image
\citep[see Fig. 10 of ][]{mahdavi05}. This asymmetry is an indication
that the core is not fully relaxed, as already suggested by the
absence of a single central galaxy. Despite this fact, hydrostatic
equilibrium beyond the inner core seems a good approximation given the values 
of \cvir\ and \mvir\ measured for this object.

{\em Abell\thin2717.} The temperature profile we have derived from
\xmm\ data declines at large radii like all of the groups in our
sample and is inconsistent with the flat profile found by
\citet{pratt05}. The origin of the difference is likely our 
improved treatment of background subtraction.
Nevertheless, our inferred $c_{200} = 4.6\pm0.2$,
$M_{200} = 1.59\pm0.06 \times 10^{14}$ \msun\ and $r_{200} =
1082\pm21$ are in good agreement with those determined both by
\citet{pratt05} and \citet{pointe05}, $c_{200} = 4.21\pm0.25$,
$M_{200} = 1.57\pm0.19 \times 10^{14}$ \msun\ and $r_{200} =
1096\pm44$. The reason for the agreement, despite the difference in
the temperature profiles, is likely the same put forward by
\citet{vikh06a}: the NFW fit implies a declining
temperature profile at large radii.

{\em RXJ\thin1159+5531.} 
Our inferred $c_{500} = 5.6\pm1.5$ is higher than the one
reported in \citet{vikh06a}, $c_{500} = 1.7\pm0.3$, using the same
\chandra\ data, though within $2.5\sigma$ given our large error bars. 
Our derived $M_{2500}=3.3\pm0.9\times10^{13}$
\msun\ and gas fraction $f_{gas,2500}=0.049\pm0.004$ are on the contrary 
in good agreement with their determination of $3.0\pm0.3\times10^{13}$
\msun\ and $0.045\pm0.002$. As for MKW 4 the key difference is in the
measured scale radius: our value of $104\pm77$, though with large
error bars, is inconsistent with their quite high value of 412 kpc,
which is again at the boundary of the radial range covered by the
data, which are of not excellent quality outside the S3 chip ($\sim
370$ kpc).

\section{Systematic errors} \label{sect_systematics}%

\def\arraystretch{1.15}
\begin{deluxetable*}{cccccccc}
\tabletypesize{\scriptsize}
\tablecaption{Systematic Error Budget \label{tab:sys}}
\tablehead{
\colhead{Group} &
\colhead{Best Fit} &
\colhead{$\Delta$Statistical} &
\colhead{$\Delta$Background} &
\colhead{$\Delta$Spectral} &
\colhead{$\Delta$Method} &
\colhead{$\Delta$Deproj} &
\colhead{$\Delta r_e$}}
\startdata
\cutinhead{$c_{\Delta}$}
NGC 5044      & 3.8  & $\pm0.1$  & -0.5  & +0.2              &  +0.6   & -0.3 ($\pm0.2$) &        \\
NGC 1550      & 4.5  & $\pm0.3$  & +0.4  & $\pm0.2$          &  -0.1   & -0.8 ($\pm0.3$) & +0.5   \\
NGC 2563      & 2.4  & $\pm1.0$  & +2.6  & +2.3              &  -0.1   & +4.5 ($\pm1.4$) &        \\
Abell 262     & 2.1  & $\pm0.2$  & +0.2  & $^{+0.8}_{-0.6}$  &  -0.4   & -0.4 ($\pm0.2$) & -0.2   \\
NGC 533       & 6.1  & $\pm0.5$  & -1.7  & -2.0              &  +1.1   & -1.5 ($\pm0.4$) & +0.9   \\
MKW 4         & 4.3  & $\pm0.3$  & -0.1  & $^{+0.3}_{-0.7}$  &  -0.3   & +0.8 ($\pm0.7$) & -0.3   \\
IC 1860       & 3.2  & $\pm0.3$  & +0.1  & $^{+0.9}_{-0.4}$  &  -1.3   &                 &        \\
NGC 5129      & 5.2  & $\pm0.9$  & +0.6  & -0.4              &  -0.3   & -0.2 ($\pm2.2$) &        \\
NGC 4325      & 2.8  & $\pm0.4$  & +0.7  & +0.9              &  +0.3   & -0.7 ($\pm0.3$) &        \\
ESO5520200    & 2.5  & $\pm0.3$  & -0.2  & -0.3              &  +0.1   & +0.2 ($\pm0.4$) &        \\
AWM 4         & 3.0  & $\pm0.3$  & +0.1  & -0.2              &  -0.1   & -0.9 ($\pm0.3$) &        \\
ESO3060170    & 2.1  & $\pm0.3$  & -0.4  & $^{+0.8}_{-0.6}$  &  -0.3   & -0.1 ($\pm0.3$) &        \\
RGH80         & 5.1  & $\pm0.5$  & +2.1  & +4.5              &  -2.6   & +2.9 ($\pm1.2$) &        \\
MS 0116.3-0115 & 2.0 & $\pm0.8$  & +0.7  & $^{+1.5}_{-0.5}$  &  +1.0   & +2.3 ($\pm1.9$) &        \\
Abell 2717    & 3.0  & $\pm0.2$  & +0.1  &  -0.2             &  -0.1   & +0.6 ($\pm0.3$) &        \\
RXJ 1159.8+5531 & 5.6 & $\pm1.5$ & -0.9  & +0.7              &  -1.2   & +2.6 ($\pm1.7$) &        \\ 
\cutinhead{$M_{\Delta}$/$10^{13}$\msun}
NGC 5044      & 1.85   & $\pm0.04$    & +0.28              & -0.10              & -0.41 & +0.34 ($\pm0.09$) &       \\
NGC 1550      & 1.42   & $\pm0.03$    & -0.04              & -0.03              & +0.02 & +0.26 ($\pm0.09$) & +0.01 \\
NGC 2563      & 0.92   & $\pm0.08$    & -0.06              & -0.17              & +0.01 & -0.24 ($\pm0.13$) &       \\
Abell 262     & 3.59   & $\pm0.14$    & -0.19              & $^{+0.24}_{-0.62}$ & +0.34 & +1.00 ($\pm0.31$) & +0.10 \\
NGC 533       & 1.30   & $\pm0.04$    & $^{+0.15}_{-0.01}$ & +0.16              & -0.04 & -0.01 ($\pm0.07$) & -0.05 \\
MKW 4         & 3.21   & $\pm0.10$    & -0.10              & $^{+0.12}_{-0.07}$ & +0.09 & -0.86 ($\pm0.18$) & +0.03 \\
IC 1860       & 2.36   & $\pm0.13$    & -0.08              & $^{+0.12}_{-0.20}$ & +0.65 &                   &       \\
NGC 5129      & 0.84   & $\pm0.07$    & $^{+0.08}_{-0.03}$ & -0.02              &       & -0.13 ($\pm0.15$) &       \\
NGC 4325      & 1.32   & $\pm0.16$    & -0.15              & -0.20              & -0.10 & +0.53 ($\pm0.45$) &       \\
ESO5520200    & 5.51   & $\pm0.51$    & +0.35              & $^{+0.70}_{-0.13}$ & -0.40 & +0.49 ($\pm0.71$) &       \\
AWM 4         & 7.38   & $\pm0.61$    & -0.27              & -0.70              & +0.16 & +2.01 ($\pm0.87$) &       \\
ESO3060170    & 5.97   & $\pm1.14$    & +1.30              & $^{+2.07}_{-0.74}$ & +0.73 & +0.68 ($\pm1.37$) &       \\
RGH80         & 1.85   & $\pm0.07$    & $^{+0.26}_{-0.14}$ & $^{+0.05}_{-0.40}$ & +0.48 & -0.07 ($\pm0.19$) &       \\
MS 0116.3-0115 & 4.92  & $\pm1.64$    & +0.46              & $^{+0.68}_{-1.42}$ & -1.12 & -0.40 ($\pm3.76$) &       \\
Abell 2717    & 10.68  & $\pm0.51$    & -0.03              & +1.02              & +0.49 & -0.76 ($\pm0.86$) &       \\
RXJ 1159.8+5531 & 6.13 & $\pm3.30$    & +0.97              & -0.29              & +0.51 & -1.87 ($\pm0.72$) &       \\ 
\cutinhead{\mstar/\lk\ (\msun/\lsun) (NFW+stars)}
NGC 1550      & 0.53   & $\pm0.20$          & +0.05              & +0.12              & +0.03  & +0.17 ($\pm0.15$) & +0.36  \\
Abell 262     & 0.54   & $\pm0.11$          & -0.13              & $^{+0.23}_{-0.37}$ & +0.12  & -0.45 ($\pm0.04$) & +0.79  \\
NGC 533       & 0.36   & $\pm0.03$          & +0.20              & +0.26              & -0.27  & +0.24 ($\pm0.04$) & -0.07  \\
MKW 4         & 0.86   & $\pm0.10$          & $^{+0.51}_{-0.66}$ & $^{+0.59}_{-0.54}$ & +0.11  & +0.15 ($\pm0.09$) & +0.60  \\
IC 1860       & 0.60   & $\pm0.14$          & +0.01              & $^{+0.02}_{-0.23}$ & +0.53  &                   &        \\
NGC 5129      & 0.06   & $^{+0.13}_{-0.06}$ & $^{+0.13}_{-0.06}$ & -0.06              & +0.06  & +0.43 ($\pm0.19$) &        \\
AWM 4         & 0.30   & $^{+0.33}_{-0.30}$ & +0.08              & $\pm0.12$          & +0.14  & +0.82 ($\pm0.38$) &        \\
RXJ 1159.8+5531 & 0.55 & $\pm0.10$          & +0.05              & $^{+0.10}_{-0.13}$ & +0.11  & -0.13 ($\pm0.13$) &        \\
\enddata
\tablecomments{The estimated error budget for each of the groups.
Excepting the statistical error, these values estimate a likely
upper bound on the sensitivity of the (best fit) value of each
parameter to various data-analysis choices and should \emph{not} be
added in quadrature with the statistical error.
The ``Best'' column indicates the best-fit value and
``$\Delta$Statistical'' the $1\sigma$ statistical error for $M_{\Delta}$ and 
$c_{\Delta}$ from Table \ref{tab:virial_pot_gas} and for \mstar/\lk\ 
from Table \ref{table_mass_to_light}. 
``$\Delta$Background'' gives the results when the
X-ray background level is set to $\pm 5\%$ of
nominal, ``$\Delta$Spectral'' gives the results when changing spectral 
analysis choices, $\Delta$Method when adopting a different approach (using 
Eq.\ref{eqn_hydrostatic_t} or Eq.\ref{eqn_hydrostatic_classic}) to mass 
modeling, $\Delta$Deproj when using projected (2D) or deprojected (3D) spectral
results (with the associated statistical error), and $\Delta r_e$ when changing the effective radius of the 
stellar profile.
\vspace*{-1\baselineskip}
} 
\end{deluxetable*}

 
In this section we address the sensitivity of our analysis 
to various systematic uncertainties and data-analysis choices which may 
impact upon our results. An estimate of the uncertainty due to these effects
for each object is given in Table~\ref{tab:sys}. The statistical error of the 
default model ($\Delta$Statistical) is also listed on the table. In the case 
of the different approach of using a deprojection tecnique ( $\Delta$Deproj), 
we also quote the corresponding ma\-gni\-tude of statistical error together with 
the associated best-fitting parameter shift in the table. 
We illustrate the effect 
of systematic errors on the best fit $c_{\Delta}$, $M_{\Delta}$ parameters,
and the stellar mass-to-light ratio \mstar/\lk.

\subsection{Background modeling and subtraction} \label{sect_systematics_bkg}
One of the most important potential sources of systematic
uncertainties in measuring the mass profiles of groups is the
background subtraction technique, in particular in the low surface
brightness regime at large radii. Our modeling technique is
particularly effective in the low temperature regime of groups, and we
take as an extreme measure to change the overall normalization of the
background model by $\pm5$\%. Such an error in the estimated
background is unlikely, but the exercise is indicative of our
sensitivity to the background.

\subsection{Spectral-fitting choices}\label{sect_systematics_spectral}
Among the variety of choices made in spectral-fitting, we explore the ones
more likely to affect to some degree the inferred gas density and temperature 
in each radial bin.

{\em The plasma code.} Different plasma codes choose from a large, overlapping, 
but incomplete set of atomic data, leading to differences in the inferred 
abundances and, therefore, density and, to a lesser extent, temperature. We 
experimented with replacing the APEC model with the MEKAL plasma model.

{\em Bandwidth.} To estimate the impact of the bandwidth on our fits,
we experimented with fitting the data with different lower limits for
the energy band. In addition to our preferred choice of 0.5 keV, we
use 0.4 keV and 0.7 keV.

{\em Hydrogen column-density.} We take into account possible deviations for
\nh\ from the value of \citet{dick90} allowing the parameter to vary by 
$\pm25$\%. 

\subsection{Deprojection method}\label{sect_systematics_depro}
\begin{figure*}[th]
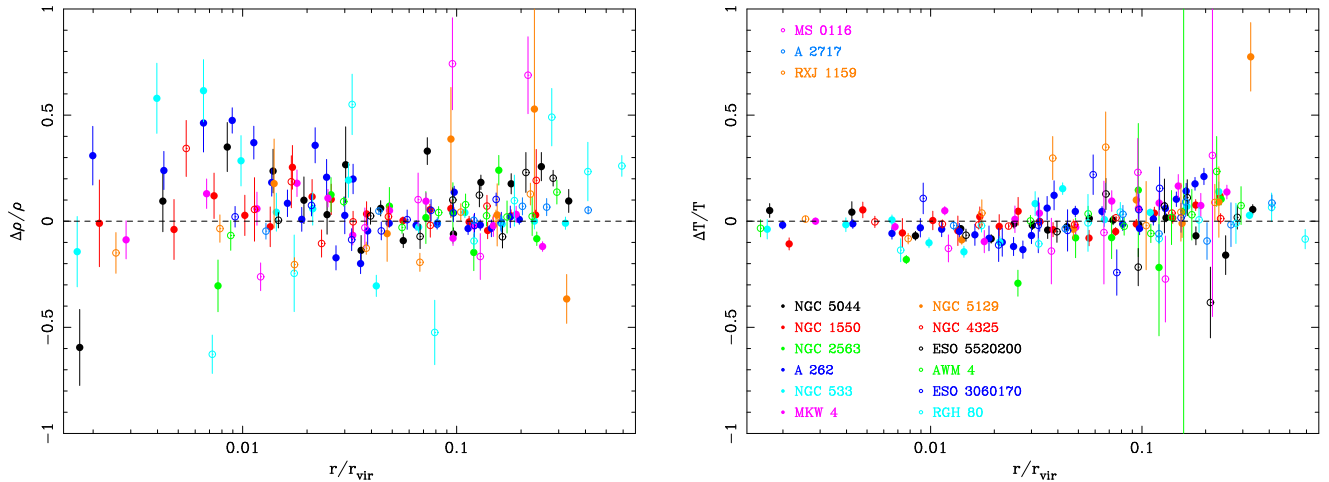

\parbox{0.5\textwidth}{
\centerline{\includegraphics[width=0.35\textwidth,angle=-90]{f7a.ps}}
}
\parbox{0.5\textwidth}{
\centerline{\includegraphics[width=0.35\textwidth,angle=-90]{f7b.ps}}
}
\caption{\label{fig.7} \footnotesize
Comparison of 2D versus 3D results: $(\rho_{2D} - \rho_{3D})/\rho_{3D}$ (left panel)
and $(T_{2D} - T_{3D})/ T_{3D}$ (right panel). Symbols used for the various systems 
are the same in both plots and they are listed on the plot in the right panel.
}
\end{figure*}

We analyzed the possible systematics involved with the projection of
3D models using instead the ``onion-peeling'' technique
\citep[e.g.,][]{fabian81,kriss83,buot00b}.  Only for the object IC 1860 we did not 
perform this exercise because of the exclusion of an inner bin (see \S
\ref{sect_objects}). The results were consistent with the ones
obtained by the 2D analysis (see Table~\ref{tab:sys}) but with larger
error bars given the quality of the current data. This is the main
reason for having adopted the 2D analysis as our default. In
Fig.\ref{fig.7} we plot as a function of the fraction of the virial
radius the quantities $\Delta\rho$/$\rho_{3D}$ (where $\Delta\rho =
\rho_{2D} - \rho_{3D}$ with $\rho_{2D}$ the value of the best fit model of the 2D
analysis) and the corresponding quantity $\Delta T$/$T_{3D}$. The
plotted errors are the fractional errors on the derived 3D
quantities. There is more scatter in the density as a consequence of
larger uncertainties in the derived 3D iron abundances, while the
temperatures determined with the two methods are generally
consistent. This fact reinforces the notion that, for the range of
temperatures spanned by the objects considered in this paper, the
spectroscopic temperatures are not biased significantly (see
discussion in \S \ref{sect_systematics_response}, \S
\ref{subsect_disc_xraymass}).

\subsection{Response weighting}\label{sect_systematics_response}
Since the effective area of the detector response of ACIS on \chandra\
and EPIC on \xmm\ are a decreasing function of energy, fitting a 1T
model to a spectrum having a range of temperature components above
$\sim 3$~keV will tend to yield a temperature that is biased low
\citep{mazzotta04,vikh06b}. This effect is negligible for average
temperatures around 1 keV \citep[see Appendix
\ref{sect_deproj_equations} and][]{buot00a} as it is the case for the
objects in our sample. As a further systematic check we applied a
straightforward averaging of the plasma emissivity over the detector
response in our method as explained in Appendix
\ref{sect_deproj_equations}. The results obtained using the response
weighting are very consistent with the ones obtained by our default
projection analysis. For example, for NGC 1550 we obtain $c_{\Delta} =
4.5\pm0.3$, $M_{\Delta}=1.45\pm0.03\times10^{13}$ \msun\ and
$M_{\star,\Delta}=11.1\pm4.0\times10^{11}$ \msun; for MKW 4 we obtain
$c_{\Delta} = 5.1\pm0.4$, $M_{\Delta}=2.91\pm0.10\times10^{13}$
\msun\ and $M_{\star,\Delta}=60.7\pm7.3\times10^{11}$ \msun; for NGC
533 we obtain $c_{\Delta} = 5.3\pm0.4$,
$M_{\Delta}=1.41\pm0.06\times10^{13}$
\msun\ and $M_{\star,\Delta}=31.6\pm2.1\times10^{11}$ \msun.

\subsection{Mass derivation method}\label{sect_systematics_method}
For each system we tried all the three methods described in \S
\ref{sect_mass}.  By using all the approaches we have an estimate of
the robustness of the inferred mass and virial quantities.  We also
include in this estimate the fact that different temperature and
density profiles may be able to fit the same data adequately but give
rise to different global halo parameters. To test this, we cycled
through each of our adopted gas density and temperature profiles.

\subsection{X-ray asymmetries and disturbances}\label{sect_systematics_asym}
There are systems displaying low level asymmetries (IC 1860),
substructure (RGH 80), and AGN cavities (A 262) or possible
AGN-induced disturbances (NGC 5044). For the objects which have a mild
degree of disturbance in the core we found that the results obtained
excluding the disturbed regions agreed with those obtained over the
entire radial range within the 1-2$\sigma$ errors. For A262 we
excluded the inner 20 kpc to avoid (1) the cavities which affect the
central 10 kpc, and (2) the stellar mass component of the central
galaxy.  In this case fitting an NFW profile gives, $c_{\Delta} =
2.4\pm0.3$ and $M_{\Delta}=3.44\pm0.15\times10^{13}$ \msun. For NGC
5044 we obtain $c_{\Delta} = 3.9\pm0.1$ and
$M_{\Delta}=1.83\pm0.04\times10^{13}$ \msun\ after excluding the
central 5 kpc where there is evidence of a disturbed morphology. We
exclude the inner 30~kpc of RGH~80 and find $c_{\Delta} = 6.5\pm1.0$
and $M_{\Delta}=1.78\pm0.08\times10^{13}$ \msun.  Finally, for IC 1860
we perform a sector analysis, extracting spectra and re-deriving our
mass profiles from suitably oriented semi-annuli, as detailed in \S
\ref{sect_objects}. We found consistent results within their $\sim
2\sigma$ errors. Therefore, we infer no systematic error associated
with including the central, mildly disturbed, regions in these systems.

\subsection{Radial Range \& Extrapolation}\label{sect_systematics_range}

It is customary to extrapolate mass profiles out to the virial radius
defined within an over-density $\Delta\sim 100-500$. This facilitates a
consistent comparison to theoretical studies which usually quote
results in this radial range, corresponding to the entire virialized
portion of the halo.  X-ray studies of global scaling relations
between mass, temperature, and luminosity also prefer to use such
large virial radii to seek the tightest relations between these global
quantities.

However, extrapolating the mass profiles can lead to systematic errors
in \cvir, \mvir, and the gas mass/fraction.  \citet{vikh06a} argue
that biased extrapolation of the gas density profiles is the main reason for 
the underestimate of gravitational masses and low normalizations of the
$M-T$ relations found with earlier X-ray telescopes
\citep[e.g.,][]{nevalainen00}, using a $\beta$ model fit for the gas
density and a polytropic approximation for the temperature
profile. \citet{rasia06} suggest that the same systematic error
affects \cvir, in the sense that a restricted radial range tends to
return a higher \cvir, in the context of the NFW profile, than the
value derived using data extending out to the virial radius.

Our procedure for mitigating extrapolation bias assuming the halo
follows an NFW profile is as follows. We obtain the mass profile
within an appropriate $r_{\Delta}$ corresponding to the outer radius
of the X-ray data for each group. The values of $c_{\Delta}$ and
$M_{\Delta}$ are extrapolated to $\Delta\approx 101$ assuming the NFW
profile applies, using the convenient approximation of
\citet{hu03}. We emphasize that we do not need to extrapolate the
models for the gas density and temperature to obtain the extrapolated
mass parameters in this manner. The most important requirement for
self-consistent extrapolation is that the NFW scale radius be
accurately measured using the available X-ray data at smaller radius.
Since our principal approach for measuring the mass profile
(parametric mass method, see \S \ref{sect_mass}) guarantees a physical
solution of the equation of hydrostatic equilibrium for an NFW DM
halo, unlike the methods used by \citet{vikh06a} and
\citet{rasia06}, and our temperature profiles are modeled with a more
sophisticated approach than the polytropic-$\beta$ model estimate, we
expect more reliable measurements of $r_s$.

The crucial factor for reliable measurement of $r_s$ is that the true
value of $r_s$ lies well within the outer radius of the X-ray data. We
illustrate this effect using those objects for which we obtained
measurements out to $\Delta=500$ (A2717, RGH 80 and RXJ 1159.8+5531).
If we exclude the outer two data points of A2717 then the new outer
data point corresponds to 320 kpc and $\Delta=2276$.  Fitting the
X-ray data over this mass range gives a best-fitting value,
$r_s=338$~kpc, uncomfortably outside the new radial range of the data
and larger than inferred using all of the data
($r_s=233$~kpc). Extrapolating this profile to $\Delta=500$ yields a
larger mass and a smaller concentration than obtained for all of the
data. Analogous results are obtained when performing this exercise for
RGH 80. For RXJ 1159.8+5531 we exclude the outer data point so that
the new outer radius is 289 kpc corresponding to $\Delta=2318$. In
contrast to A2717 and RGH 80, when fitting over this smaller radial
range, we obtain a scale radius $131\pm76$ kpc, still well within the
outer radius. The derived mass parameters are, $c_{2500}=2.2\pm0.6$
and $M_{2500}=3.69\pm0.86\times10^{13}$ \msun.  Extrapolating these
values to $\Delta=500$ we obtain $c_{500}=4.7\pm1.1$,
$M_{500}=7.10\pm3.25\times10^{13}$ \msun, in excellent agreement with
the results obtained over the whole data range presented in Table
\ref{tab:virial_pot_gas}.

This exercise suggests that measurements of $c_{\Delta}$ and
$M_{\Delta}$ should be reliable provided the NFW scale radius lies
well within the outermost radius covered by the data, as is the case
for all the objects in our sample. Agreement with the optical
determination of the scale radius for the two objects in common with
\citet{rines06} adds further strength to the results (see \S
\ref{sect_objects}).

Unfortunately, extrapolation of the gas mass and gas fraction is less
reliable.  If we extrapolate our models out to a virial radius
corresponding to $\Delta\sim 101$ we obtain gas fractions consistent
with the cosmic value in 12 of 16 cases. In 4 cases the extrapolated
gas fractions exceed the cosmic value derived by
\emph{WMAP}, suggesting a problem with the extrapolation.  
All these systems possess a flat slope of the gas density profile (
$\beta < 0.5$) at the edge of the data range.
This type of behaviour has been noted previously by simulations and simple analytic
models which pointed out how the $\beta$ model overestimates gas mass
(and underestimates gravitational masses based on $\beta$ model fits),
because it returns a biased low $\beta$ due to the restricted range of
radii where the fit is performed
\citep{navarro95,bartelmann96,borgani04,komatsu01}. Indeed,
\citet{vikh06a} finds evidence for a steepening of the gas density 
slope with radius in clusters.

\subsection{The stellar mass profile of the central galaxy}
\label{reff}
To account for the stellar component we adopted a De Vaucoleurs model with 
effective radius being fixed to that determined by \twomass. 
The derived stellar mass is most sensitive to the effective radius.
The difference in effective radii measured in different optical bands, as
evident in Table~\ref{tab:optical} is mainly due to the use of different fitting 
ranges/sensitivity \citep[e.g.,][]{fisher95} and to a radial color 
gradient, reflecting gradients in the metallicity or age of the stellar 
population \citep[e.g.,][]{pahre99}. 
Though the true stellar mass is more reliably 
determined from K-band data, we investigated the sensitivity of our
parameters, in particular the value for \mstar/L, to the choice of
$r_e$, by replacing the K-band $r_e$ for each galaxy with the larger
B-band value, listed in the Third Reference Catalog of Bright Galaxies
\citep[RC3:][]{rc3}.
The stellar mass, and consequently \mstar/L, increases systematically, 
with the only exception being NGC 533, when using the larger B-band effective
radius. But, importantly, the concentration and mass are not affected, showing
that the main conclusions of our paper regarding these two quantities are not 
sensitively dependent on the adopted stellar template.

We studied the possible contribution to the stellar mass of non
central galaxies within $r_s$ using known member galaxies with
\twomass\ photometry as listed in \ned. For most of the objects they
contribute no more than 10\% of the total light. This is consistent
with the more general result of \citet{linmohr04} who showed that the
ratio of BCG-to-total galaxy light decreases with increasing cluster
mass indicating that 30-50\% of the total light in galaxies is in the
BCG for group-scale systems ($10^{13}$ \msun $< M_{200} < 10^{14}$
\msun). Notable exceptions are infact the most massive 
objects in our sample: A262 has a 77\% and AWM 4 a 39\% additional 
contribution from non central galaxies within $r_s$.

\section{Discussion} \label{sect_discussion}

\subsection{\cvir-\mvir\ relation}\label{c_M}

 \begin{figure*}[t]
\centerline{\includegraphics[width=0.6\textwidth,angle=-90]{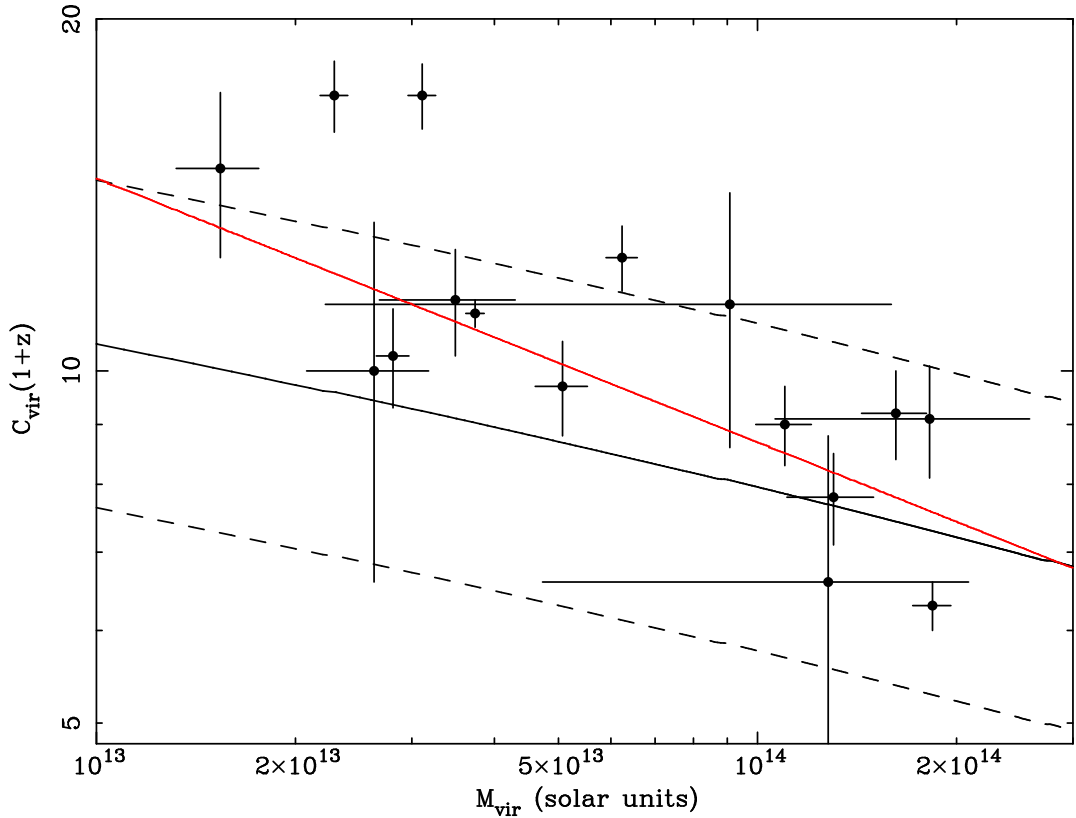}}
\caption{\label{fig.8} \footnotesize
Concentration parameters \cvir\ (multiplied by 1+z) versus the groups 
mass \mvir.
The solid black line represents the median \cvir(\mvir) relation and the outer 
dotted lines the $1\sigma$ scatter ($\Delta$log\cvir $=0.14$) calculated 
according to the model of \citet{bullock01} with parameters F=0.001 and K=3.12
and for a concordance model with $\sigma_{8} =0.9$. The solid red line represents the best fitting power law relation discussed in the text. 
}
\end{figure*}

In Fig. \ref{fig.8} we plot the \cvir\--\mvir\ relation fits to the
\xmm\ and \chandra\ data ($\Delta\approx 101$, see Table
\ref{tab:delta}). To obtain an empirical description of the relation we
fitted a simple power-law model following the approach described in
\citet{buote06b}. That is, we fitted the data with a linear relation
of the form ${\rm{log}}\, (1+z)c_{\rm{vir}} =
\alpha\,{\rm{log}}[M_{\rm{vir}}]+b$ using the BCES estimator of
\citet{akritas96} with bootstrap resampling.  
We obtain $\alpha=-0.226\pm0.076$, implying that the concentration decreases
with increasing mass at the 3$\sigma$ level. The previous studies of
clusters ($>10^{14}$~\msun) with \chandra\ and \xmm\ found
$\alpha\approx 0$ -- very consistent with a constant
\cvir\--\mvir\ relation \citep{pointe05,vikh06a}. 
Therefore, it is the lower mass range, $10^{13}-10^{14}$ \msun,
appropriate for groups that provides crucial evidence that \cvir\
decreases with increasing \mvir\ as expected in CDM models
\citep[e.g.,][]{bullock01}.

The best-fitting power-law model is plotted in Fig. \ref{fig.5} along
with the theoretical prediction of the $\Lambda$CDM obtained using the
model of \citet{bullock01} with parameters $(F=0.001,K=3.12)$ intended
to represent halos up to masses $\sim 10^{14}$~\msun. Also shown is the
predicted $1\sigma$ intrinsic scatter for the $\Lambda$CDM model. For
$M\ga 4\times 10^{13}$~\msun the $\Lambda$CDM model is a good
representation of the X-ray data. For lower masses, the observed
\cvir\--\mvir\ appear to exceed the prediction. Allowing for a
$\approx 10\%$ increase in the concentrations predicted by the
$\Lambda$CDM model for the most relaxed, early forming halos
\citep{jing00,wechsler02,maccio06} helps to bring the model into
better agreement with the observations.

We infer an intrinsic scatter, $0.03\pm 0.02$, in $\log_{10}\, (1+z)c$
for the empirical power law relation \citep[see][]{buote06b} which
is considerably less than the value of $\approx 0.14$ obtained for
$\Lambda$CDM halos \citep{jing00,wechsler02,maccio06}. For the most
relaxed, early forming halos $\Lambda$CDM simulations typically find a
smaller scatter $\sim 0.10$ \citep{wechsler02,maccio06}, though the
most relaxed halos studied by \cite{jing00} have a scatter of 0.07
(after converting between $\ln$ and $\log_{10}$ with a factor of
2.3). The small scatter we have measured about the power-law relation
only agrees with CDM simulations if these halos are the most relaxed,
early forming systems.

The data-model comparison has been made for a $\Lambda$CDM model with
parameters obtained from the first year WMAP results
\citep{spergel03}, in particular with \omegam=0.3 and
$\sigma_{8}=0.9$. With the lower values favored by the three year data
release of WMAP \citep{spergel06} the predicted concentrations are lower 
\citep[e.g., see discussion in][]{maccio06}. The implications are
discussed in \citet{buote06b}.

\subsection{The detection of central stellar mass}

A good fit of the NFW profile to the total gravitating matter of relaxed, 
$T >$ 3 keV, massive clusters, without any significant deviation arising
from the central stellar mass, appears to be a common feature of X-ray studies 
\citep{pointe05,vikh06a,zappacosta06}. On the contrary, relaxed 
bright elliptical galaxies always require a dominant contribution of
stellar mass \citep{humphrey06a}. The intermediate mass range explored
here shows a mixed behavior: some low-mass, group-scale objects and
three poor clusters (AWM 4, RXJ 1159 and Abell 262) do show evidence
of stellar mass , while there are examples of objects whose
gravitating mass profile is described by just NFW.  (Note that RXJ
1159, A 262 and MKW 4 were also shown to have an excess core mass
profile above that indicated by NFW in the analysis of
\citealt{vikh06a}.)

An important issue emerging in the analysis is how well the two key
components, the stellar component associated with the central galaxy
and the DM, are sampled by the X-ray data. It is expected that the
stellar component is most relevant within the inner 10-20~kpc while
the DM should dominate the mass budget elsewhere. To reveal and
measure adequately the stellar mass, enough density and temperature
data points are required in the inner $\sim 20$~kpc, depending as well
on the amount of stellar mass present (implied by \lk).  The omission
of the stellar component in the mass modeling, proposed as a possible
source of abnormally high $c$ \citep{mamon05a}, is certainly a factor
for relatively nearby objects with data densely sampling the inner
dominated stellar core but not extending to large radii, as for the
objects analyzed with \chandra\ data in \citet{humphrey06a}. The
effect is less pronounced in the objects analyzed in this sample,
where the data extend to large enough radii, but with comparatively
less density of data points in the inner 20 kpc, in particular for
objects with only \xmm\ observations. The presence of data at large
radii prevent to obtain large values of \cvir ($\geq 30$) when
fitting the wrong NFW model to objects which require stellar mass.

For the objects which require stellar mass and have 2-3 data bins in the 
inner 20 kpc (NGC 1550, A262, NGC 533, MKW 4) the derived stellar M/L ratios 
are consistent with the range of values found in our analysis of a sample of 
elliptical galaxies \citep{humphrey06a}: the (unweighted) mean M/L ratio of 
these four objects is 
$0.57\pm0.21$ which is consistent within $1\sigma$ with $0.76\pm0.24$, the 
mean stellar M/L ratio of the objects in \citet{humphrey06a}, though
a 25\% difference is present. This result reinforces the relevance of the 
sampling of the inner region: on average the objects in \citet{humphrey06a}
have $\sim 7$ data bins in the inner 20 kpc allowing a more accurate 
measurement of the stellar mass.
The measures for objects with low
resolution \xmm\ observations are likely biased low, as we determined
for objects with both \xmm\ and \chandra\ observations. 

For a single
burst stellar population (SSP) with age ranging from 9-13 Gyr and
metallicity ranging from 0.5-2 solar, the K-band stellar mass-to-light
ratio is expected to take values in the range 0.86-1.16 for a Kroupa
\citep{kroupa01} IMF and 1.28-1.49 for a Salpeter IMF
\citep{salpeter55}. IMF (We have used linearly interpolated synthetic
\mstar/\lk\ values based on the stellar population models of
\citet{maraston98} from updated model-grids made available by the
author\footnote{http://www-astro.physics.ox.ac.uk/~maraston/Claudia's\_Stellar\_Population\_Models.html}
and converting from their definition of $M_{Ks}\odot = 3.41$).
The measurements are, therefore, in reasonable agreement with the SSP
models assuming a Kroupa IMF, given the uncertainties in both the data
and models.

Clearly the X-ray determination of the stellar mass contribution in
these objects can benefit from deeper observations, and the
systematics involved in the modeling of the stellar profile, like the
value of the effective radius, impact the results
considerably. However, the excellent agreement between the gravitating
mass-to-light ratio at the effective radius obtained from X-rays and
globular cluster kinematics for the elliptical galaxy NGC 4649 by
\citet{bridges06} provides strong support for the reliability of the
stellar mass-to-light ratio determined from X-rays in that system.

Stellar mass has not been detected for NGC 5044, RGH 80 and NGC 4325, systems 
which do have \chandra\ data allowing a reasonable sampling of the inner core.
Better \chandra\ data would be required in the core of NGC 5044 to measure 
possible localized disturbances due to AGN activity, which can be a likely 
source of systematics in the mass measurement in the inner 10-20 kpc.
Hint of disturbances (cavities) due to AGN activity have also been detected 
in NGC 4325 \citep{russel07}.
There is evidence from the optical and the \chandra\ X-ray image that the core 
of RGH 80 may not be completely relaxed. Therefore localized departures 
form hydrostatic equilibrium in the core of these systems are likely 
explanations for the failure to detect stellar mass. 

If we allowed the DM profile to be modified by adiabatic contraction,
we obtained substantially smaller \mstar/\lk\ for our data, which are
more discrepant with SSP models, casting doubt on the importance of
the adiabatic contraction process. We obtained similar results in our
study on elliptical galaxies \citep[][and discussion therein of other 
observational results]{humphrey06a}. Recently \citet{gustafsson06} 
proposed that also the details of the feedback and its effect 
upon the concentration of the baryons is an important ingredient for 
the determination of the final contracted dark matter halo.

We tested the sensitivity of the measured concentrations and gravitational 
masses to the stellar template by substituting the effective radius of the
De Vaucoleurs model derived by \twomass\ with the larger value listed in RC3.
By doing that the stellar mass-to-light ratio can increase up to 80\% but
the concentration and gravitational mass are not affected as dramatically 
(see \S \ref{reff}). The latter measurements seem therefore robust to
the assessment of the precise best fit model for the light profile of BCGs and the 
precise measurement of their size 
\citep[a task which requires particular care, e.g.,][and it is beyond the scope of this paper]{bernardi07,seigar06}.

\subsection{Gas fractions}\label{f_gas}

\begin{figure*}[t]
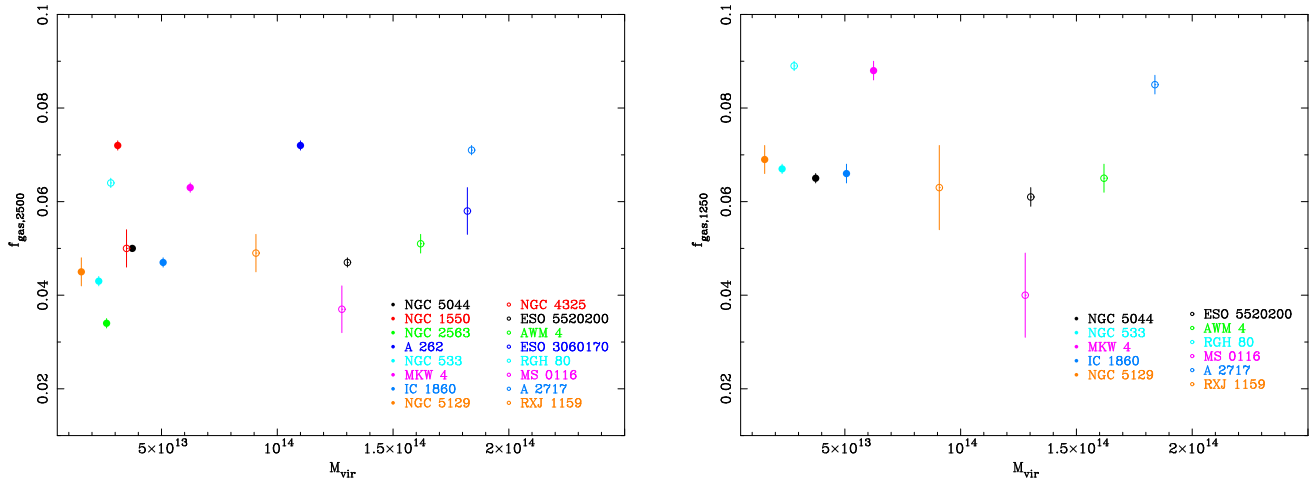

\parbox{0.5\textwidth}{
\centerline{\includegraphics[width=0.35\textwidth,angle=-90]{f9a.ps}}
}
\parbox{0.5\textwidth}{
\centerline{\includegraphics[width=0.35\textwidth,angle=-90]{f9b.ps}}
}
\caption{\label{fig.9}
Observed gas fractions within $r_{2500}$ (left panel) and within $r_{1250}$ (right panel) as a function of the virial mass. Errors bars on the virial mass have not been shown for clarity of the plot.
}
\end{figure*}

In Fig. \ref{fig.9} we plot the gas fractions for the objects in our
sample, calculated within a radius corresponding to an over-density of
$\Delta=2500$ and $\Delta=1250$, as a function of virial mass.  No
significant trend of gas fraction with mass is present, while there is
significant object-to-object scatter: the average value of the gas
fractions are, $f_{gas,2500}=0.053\pm0.012$ and
$f_{gas,1250}=0.069\pm0.014$.  The mean value of $f_{gas,2500}$
obtained for the groups in our sample is significantly smaller than
that obtained from the hot, massive clusters ($T > 5$ keV) studied by
\citet{alle04a}, $f_{gas,2500}=0.118\pm0.016$, and \citet{vikh06a},
$f_{gas,2500}=0.092\pm0.004$. (Note we quote the mean and standard
deviation, not the gaussian error-weighted mean and error, because of
the possibility of non-gaussian contributions to the gas fraction
distributions, such as intrinsic scatter caused by scale-dependent
feedback processes.) The fractional error obtained for our groups,
$\sigma_f/f=0.2$, exceeds the values of 0.14 and 0.04 for,
respectively the clusters of \citet{alle04a} and
\citet{vikh06a}. Therefore, there is a clear mass dependence on the
gas fraction (mean and fractional scatter), not surprising given that
the expected feedback energy injection by AGN should be more severe at
the group scale.

The extrapolation of gas quantities outside the radial range of the
data is dangerous (see \S \ref{sect_systematics_range}).  However, for
most of the groups in our sample, we find that the extrapolated gas
fraction, coupled with the estimate of the stellar mass, yield global
baryon fractions consistent with the universal value; i.e., consistent
with notion that X-ray bright groups are baryonically closed
\citep{mathews05}. This result suggests that for the objects in our
sample for which the slope of the gas density profile is not too flat 
($\beta \gtrsim 0.5$), the
extrapolation of the gas fraction is also fairly reliable. Data at
large radii are much needed to further explore this issue.

\subsection{Temperature profiles}\label{t_prof}

The temperature profiles of our groups (Figures \ref{fig.3},
\ref{fig.4}, \ref{fig.5} and \ref{fig.6}) exhibit the same
behavior characteristic of cool core clusters
\citep{mark98a,degr02,piffa05,vikh05a,pratt06b}; i.e., the temperature
profile rises outwards from the center, reaches a maximum, and then
falls at large radius. To examine the self-similarity of the profiles,
we first rescaled them in terms of the virial radii. Then we
normalized each profile according to the gas-mass-weighted temperature
($T_{gmw}$) computed between 0.1-0.3 \rvir\ using the temperature and gas
density models we derived for each system.  (Gas-mass weighting should
be more closely related to the gravitational potential than emission
weighting, e.g., \citealt{mathiesen01}.) We find that the scaled
temperature profiles are approximately self-similar 
for $r > 0.15$\rvir, but there is a large amount of scatter at smaller radii 
($r < 0.1$\rvir). This behavior is qualitatively similar to that found in
clusters by \citet{vikh06a}, though the large scatter in the cores
suggest that feedback (AGN) processes have had a more dramatic impact
at the group scale; e.g., the striking isothermal core of AWM 4.

Unlike the rise-then-fall temperature profiles observed for relaxed,
cool core groups and clusters with $M\ga 10^{13}$~\msun, the three
galaxy scale systems ($M<10^{13}$~\msun) we studied in
\citet{humphrey06a} all have temperature profiles that decrease
monotonically with increasing radius.  Hence, the temperature
profiles observed in our present study provide further support for the
suggestion we made in
\citet{humphrey06a} that $\approx 10^{13}$~\msun\ represents the mass-scale
demarcating the transition between (field) galaxies and groups.  The
dramatic change in M/L ratios observed at this mass scale from optical
and lensing studies \citep[][and references therein]{parker05} gives
additional evidence that $\approx 10^{13}$~\msun\ is a special mass
scale.

\subsection{Reliability of X-ray mass and concentration estimates}\label{subsect_disc_xraymass}

Key sources of systematic errors in the X-ray determination of mass
and concentration parameters discussed in the literature can be listed
as: the applicability of hydrostatic equilibrium, a correct
interpretation of the temperature measured by X-ray satellites
\citep[i.e., spectroscopic versus emission weighted, e.g.][]{mazzotta04}
and the restricted radial range over which the mass is inferred.

The results presented in this paper show how mass constraints for
X-ray bright groups/poor clusters derived from good quality \xmm\ and
\chandra\ observations can be of the same quality as obtained for hot,
massive clusters.  The objects in our sample have been chosen
following similar criteria for selecting relaxed clusters for mass
studies \citep[e.g.,][]{pointe05,vikh06a}. Indeed, our results show
strong support for a scenario where hydrostatic equilibrium is an
excellent approximation. The mass profiles inferred from density and
temperature profiles are in good agreement with the predicted
quasi-universal NFW profile, and the concentration parameters are as
expected (i.e., for $\Lambda$CDM) for the masses of these objects. The
observed trend toward more concentrated halos, as expected from
numerical simulations for relaxed halos which have not experienced a
recent major merger, provides further verification our selection
criteria.  The fits based on the equation of hydrostatic equilibrium
were able to model complicated temperature and density profiles
(assuming a simple NFW DM halo and central stellar component) which
would be surprising if the gas is significantly out of hydrostatic
equilibrium.

The estimate of the real temperature from the projected X-ray
temperature has been discussed as a source of systematic error in
X-ray mass estimates.  If a spectrum contains several components with
different T and metallicity the ``spectroscopic temperature'' derived
from a single temperature fit is biased toward lower temperature
components \citep{mazzotta04,vikh06a}.  But the particular temperature
range of 1-3 keV explored in this paper is sensitive to the presence
of complex thermal structure, because the Fe-L shell lines complex is
very prominent in the spectrum and able to discern different
temperature components \citep{buot00a,buot00b,bohr02}. If a spectrum
contains components with different temperatures, the residuals have a
characteristic shape, originally noted in \asca\ large beam spectra as
the `Fe bias'', because the Fe abundance derived from such a spectrum
is biased low \citep{buot00a}. Although the Fe abundance is biased low
in such cases, the inferred average temperature is not biased for
$\sim 1$~keV systems, as shown in this paper and in Appendix of
\citet{buot00a}.  (Note that such an underestimate in the Fe abundance
will lead to an overestimate of the gas density. But if the bandpass
extends down at least to $\approx 0.5$~keV, as it does for the ACIS
and EPIC data, the Fe bias is reduced significantly
\citep{buot00c}. Further reduction in the Fe bias from previous
large-beam \asca\ studies results from using much thinner annuli,
especially in the core regions exhibiting the steepest temperature
gradients where the bias would be largest.  Considering that the Fe-L
line emission contributes less than the bremsstrahlung over the ACIS
and EPIC bandpasses, and the gas density varies as the square root of
the emissivity, the reduced Fe Bias leads to a minor overestimate of
the gas density that is less than the statistical error and other
systematic errors considered in this paper.)

What remains to be determined is whether the gas is truly
single-phase. The use of annuli with the smallest possible radial
width, allowed by the superb \chandra\ spatial resolution, is crucial
to ensure that the multiphase appearance is not simply due to a single
phase temperature gradient. The multiphase appearance of
\xmm\ spectra extracted in the core of groups like NGC 533
\citep{kaastra04} and RGH 80 \citep{xue04} is caused mainly by this
reason.  This possible error, i.e., assuming the gas is single-phase
when it is actually multiphase, only could affect the innermost
regions where the temperature gradient is most pronounced.  Therefore
this possible source of systematic in some systems can be relevant
only for obtaining the most precise measurement of the stellar mass,
but it is unimportant for determinations of the halo mass and
concentration.

For the objects in our sample the scale radius is well within the
radial range covered by the data. Therefore a restricted radial range,
even for objects for which we reach an over-density of 2500, should not
be an important source of systematic error in the measurement of
concentration parameters, as we showed in \S
\ref{sect_systematics_range}.

We also found our results are not sensitive to mild disturbances
related to the presence of a central AGN or not fully relaxed
dynamical state (\S \ref{sect_systematics_asym}).  The derived
concentrations and masses are also quite insensitive to errors in the
shape of the stellar mass profile of the central galaxy, (\S
\ref{reff}).

It is still very interesting and desirable to obtain data at larger radii 
with offset observations performed by the current 
generation of X-ray observatories, in particular \xmm, rather than still rely 
on \rosat, in particular to measure gas masses and gas fraction (and to 
further constrain the total mass). \citet{mole04a}, addressing 
the issue of our ignorance of the outer regions of hot massive clusters, 
pointed out how a combination of reduction of particle background to lower 
levels compared to the cosmic background and the use of a differential measure,
to improve the knowledge of the actual background in the observation, are key 
to a successful measure of the very low surface brightness regime at large 
radii. The discussion in \citet{mole04a} is focused in particularly in 
measuring the exponential cut-off of the bremsstrahlung spectrum of a 
cluster with $T >$ 3 keV. For
the particular case of groups/poor clusters, where instead the temperature is 
determined by the Fe-L shell, the clear separation of the source component 
(at least over the radial range where the temperature is not declining at 
values comparable to the temperature of the soft Galactic background) from all 
the other background components is an effective way of making a 
differential measure, because we know both the source and the background. 
The use of the improved 
\xmm\ capabilities in term of collecting area and spectral resolution respect 
to \rosat\ will also lift the likely important metallicity-density degeneracy 
(which is even more important at large radii, being the group emission due to 
line emission).
Furthermore the planned future X-ray observatories, like \emph{Xeus} and 
\emph{Constellation-X} will have smaller field of view and the mapping of the 
outer regions of nearby systems will be even more demanding in terms of 
observing time.

 \section{Summary} \label{sect_conclusions}

Using \chandra\ and \xmm\ data we have obtained detailed density,
temperature and mass profiles of 16 groups/poor clusters which were
selected to be highly relaxed systems with the best available data. In
summary:

\begin{enumerate}

\item The mass profiles were well described by a  two component model: an NFW 
model for the DM and a De Vaucoleurs stellar mass model for 8 objects. For 
objects without adequate sampling in the inner 20 kpc and for NGC 5044, 
NGC 4325 and RGH 80 a pure NFW model was a good fit of the data.
A possible explanation for the failure to detect stellar mass in NGC 5044 and NGC 4325 
is localized disturbance by AGN activity and for RGH 80 not complete relaxation 
in the core.
For objects with evidence of stellar mass, the stellar mass-to-light ratio in 
the K band was found to be in approximate agreement with simple 
stellar population synthesis models, assuming a Kroupa IMF.

\item Adopting more complicated models, like introducing adiabatic 
contraction or the recently proposed N04 DM profile did not improve
the fits.  With the available data AC produces too low stellar
mass-to-light ratios and N04 has too high inverse Sersic indexes.

\item The measured \cvir-\mvir\ relation agrees
with the predictions of $\Lambda$CDM with $\sigma_{8}=0.9$ and
\omegam=0.3. In particular in the mass range of our group sample the
expected decrease of \cvir\ with \mvir\ has been detected for the
first time. There is a trend, common to all X-ray observations, toward
more concentrated halos, which can be understood in terms of a
selection bias, already explored in numerical simulations, toward
relaxed, earlier forming systems.

\item The gas fraction measured at an over-density of 2500 is lower than the 
one measured for hot massive, clusters, and has higher scatter, as
expected if feedback has played a more severe role at this mass
scale. However the gas fractions increase with radius, and for objects
with data extending to large radii, these objects are consistent with
being baryonically closed. However the gas fractions increase with radius, and 
for objects with data extending to large radii, these objects are consistent with
being baryonically closed.\\

\item When rescaling the radial temperature profiles in terms of
\rvir\ and also the gas-mass-weighted temperature (evaluated over 0.1-0.3~\rvir), we
find the scaled profiles show a fair amount of similarity beyond 0.15
\rvir. In the core ($r < 0.10$ \rvir) the scaled profiles have a large
amount of scatter, again suggesting the important role of feedback in
these groups.

\item We tested the robustness of our results performing a careful
analysis of possible systematic errors, like background subtraction, 
departures from hydrostatic equilibrium, deprojection method
and found none of them to seriously affect our analysis and results.

\end{enumerate}

\begin{acknowledgements}
We thank the referee for helpful suggestions which improve the presentation
of the paper. 
We would like to thank Oleg Gnedin for making available his adiabatic
compression code. We thank A. De Luca, S. Ettori, A. Kravtsov, G. Mamon,
P. Mazzotta, S. Molendi and A. Vikhlinin for interesting discussions.  This work is
based on observations obtained with \xmm\, an ESA science mission with
instruments and contributions directly funded by ESA member states and
the USA (NASA).  This research has made use of data obtained from the
High Energy Astrophysics Science Archive Research Center (HEASARC),
provided by NASA's Goddard Space Flight Center.  This research has
also made use of the NASA/IPAC Extragalactic Database (\ned) which is
operated by the Jet Propulsion Laboratory, California Institute of
Technology, under contract with NASA.  In addition, this work also
made use of the HyperLEDA database.
D.A.B., and F.G.\ gratefully acknowledge partial support from NASA
grant NAG5-13059, issued through the Office of Space Science
Astrophysics Data Program.  Partial support for this work was also
provided by NASA through Chandra Award Numbers GO4-5139X and GO6-7118X
issued by the Chandra X-ray Observatory Center, which is operated by
the Smithsonian Astrophysical Observatory for and on behalf of NASA
under contract NAS8-03060. We also are grateful for partial support
from NASA-XMM grants, NAG5-13643, NAG5-13693, and NNG05GL02G.
\end{acknowledgements}

\appendix
\section{A: Masses and concentration parameters at different over-densities} \label{sect_overdensities}
Following \citet{arna05} and \citet{vikh06a} we quote in Table \ref{tab:delta}
 mass, concentration parameters and characteristic radii at different 
over-densities. The virial 
quantities obtained in our fits have been rescaled to the often used 
over-density levels of $\Delta =$ 500, 200 and  $\Delta_{\rm{vir}}$ listed in 
 Table \ref{table_obs} using the fitting formula of \citet{hu03}, which for 
$c <$ 20 is accurate to 0.3\%.

\def\arraystretch{1.35}
\begin{deluxetable*}{llllllllll}
\tabletypesize{\scriptsize}
\tablecaption{Results for the NFW profile at different over-densities \label{tab:delta}}
\tablehead{
\colhead{Group} &
\colhead{\cvir} &
\colhead{\rvir} &
\colhead{\mvir} &  
\colhead{$c_{500}$} &
\colhead{$r_{500}$} &
\colhead{$M_{500}$} &
\colhead{$c_{200}$} &
\colhead{$r_{200}$} &
\colhead{$M_{200}$} \\
\colhead{(kpc)} & & \colhead{(kpc)} & \colhead{($10^{13}\,M_\odot$)} &  & \colhead{(kpc)}  &\colhead{($10^{13}\,M_\odot$)} &  & \colhead{(kpc)}  &\colhead{($10^{13}\,M_\odot$)} }
\startdata
NGC 5044  & $11.1\pm0.3$ & $860\pm9$ & $3.74\pm0.12$ & $5.7\pm0.1$ & $443\pm4$ & $2.51\pm0.07$ & $8.4\pm0.2$ & $653\pm7$ & $3.21\pm0.10$ \\
NGC 1550  & $17.0\pm1.1$ & $811\pm13$ & $3.11\pm0.15$ & $9.0\pm0.6$ & $428\pm6$ & $2.24\pm0.09$  & $13.0\pm0.9$ & $621\pm9$ & $2.73\pm0.12$\\
NGC 2563  & $9.9\pm3.4$ & $762\pm55$ & $2.63\pm0.55$ & $5.1\pm1.9$ & $390\pm22$ & $1.72\pm0.28$ & $7.5\pm2.6$ & $577\pm39$ & $2.24\pm0.43$ \\
Abell 262  & $8.9\pm0.7$ & $1232\pm38$ & $11.00\pm1.06$ & $4.5\pm0.4$ & $624\pm15$ & $7.02\pm0.52$ & $6.7\pm0.5$ & $930\pm27$ & $9.29\pm0.82$ \\
NGC 533   & $16.9\pm1.2$ & $727\pm12$ & $2.29\pm0.11$ & $9.0\pm0.7$ & $385\pm5$ & $1.65\pm0.06$ & $13.0\pm0.9$ & $559\pm8$ & $2.02\pm0.09$ \\
MKW4      & $12.3\pm0.8$ & $1012\pm18$ & $6.24\pm0.34$ & $6.4\pm0.5$ & $527\pm8$ & $4.27\pm0.18$ & $9.4\pm0.7$ & $773\pm13$ & $5.40\pm0.27$ \\
IC 1860   & $9.5\pm0.9$ & $946\pm29$ & $5.07\pm0.46$ & $4.9\pm0.5$ & $484\pm11$ & $3.31\pm0.23$ & $7.2\pm0.7$ & $718\pm20$ & $4.30\pm0.36$ \\
NGC 5129  & $14.6\pm2.4$ & $636\pm29$ & $1.54\pm0.22$ & $7.7\pm1.3$ & $335\pm12$ & $1.09\pm0.12$ & $11.2\pm1.9$ & $488\pm21$ & $1.35\pm0.18$ \\
NGC 4325  & $11.2\pm1.2$ & $833\pm57$ & $3.49\pm0.81$ & $5.8\pm0.7$ & $432\pm25$ & $2.36\pm0.45$ & $8.6\pm1.0$ & $635\pm41$ & $3.01\pm0.65$ \\
ESO5520200  & $7.6\pm0.7$ & $1288\pm61$ & $13.03\pm1.95$ & $3.9\pm0.4$ & $650\pm24$ & $8.05\pm0.94$ & $5.8\pm0.6$ & $976\pm43$ & $10.90\pm1.51$ \\
AWM 4      & $8.9\pm0.8$ & $1384\pm53$ & $16.19\pm1.82$ & $4.6\pm0.5$ & $708\pm23$ & $10.44\pm0.99$ & $6.8\pm0.6$ & $1054\pm38$ & $13.75\pm1.46$ \\
ESO3060170  & $8.8\pm1.0$ & $1436\pm141$ & $18.21\pm7.58$ & $4.5\pm0.6$ & $733\pm59$ & $11.67\pm3.72$ & $6.7\pm0.8$ & $1093\pm100$ & $15.44\pm5.86$ \\
RGH 80   & $9.9\pm1.0$ & $771\pm15$ & $2.81\pm0.16$ & $5.1\pm0.5$ & $398\pm5$ & $1.86\pm0.07$ & $7.6\pm0.8$ & $588\pm10$ & $2.41\pm0.12$ \\
MS 0116.3-0115  & $6.3\pm2.1$ & $1272\pm220$ & $12.79\pm8.06$ & $3.1\pm1.2$ & $634\pm85$ & $7.54\pm3.41$ & $4.8\pm1.7$ & $962\pm153$ & $10.55\pm5.92$ \\
Abell 2717  & $6.0\pm0.3$ & $1432\pm31$ & $18.39\pm1.22$ & $3.0\pm0.2$ & $710\pm12$ & $10.68\pm0.52$ & $4.6\pm0.3$ & $1082\pm21$ & $15.10\pm0.89$ \\
RXJ 1159.8+5531  & $10.6\pm2.6$ & $1110\pm177$ & $9.08\pm6.86$ & $5.6\pm1.5$ & $585\pm73$ & $6.16\pm3.29$ & $8.3\pm2.1$ & $861\pm127$ & $7.87\pm5.33$ \\
\enddata
\end{deluxetable*}
\def\arraystretch{1.0}

\clearpage
\section{B: Relating X-Ray Observations to Spherical Models of Coronal
Gas with Variable Plasma Emissivity} \label{sect_deproj_equations}  

We relate the spherical models of density ($\rho_g$) and temperature
($T$) of the hot gas in elliptical galaxies, galaxy groups, and
clusters (\S \ref{sect_mass}) to the parameters obtained from
conventional spectral fitting of X-ray data, such as from
\chandra\ and \xmm.  Since much of the relevant material is scattered
throughout the literature spanning over at least 30 years, this
appendix provides an opportunity to give a self-contained, up-to-date
presentation. Our treatment fully accounts for radial variations of
the plasma emissivity which are often neglected, especially when
inferring the gas density.  We follow the standard procedure where
emission from coronal plasma characterized by a single temperature
(1T) is fitted to the X-ray spectrum extracted from a circular annulus
(2D) or, if the data have been deprojected, a spherical shell
(3D). The density and temperature (and abundance) parameters obtained
from such a 1T fit are compared to the emission-weighted (and
projected if 2D) spherical models of $\rho_g$ and $T$ (and
abundances). It is assumed the annuli are chosen to be sufficiently
wide so that the detector point spread function may be neglected.  We
discuss additional weighting by detector responses at the end of this
section.

For any spectral quantity, such as luminosity (erg s$^{-1}$), that
results from integrating between photon energy $E_1$ to energy $E_2$,
we use the notation,
\begin{equation}
L(\Delta E) \equiv \int_{E_1}^{E_2} \frac{ {\rm d}L }{ {\rm d}E } {\rm d}E,
\end{equation}
to represent the bandpass integration. For a coronal plasma emitting
within a volume ${\rm V}_i$ at a temperature $T_i$ with chemical
abundances $Z_{{\rm Fe},i}$, $Z_{{\rm O},i}$, $Z_{{\rm Si},i}$, etc.\
the luminosity is given in {\sc xspec} by,
\begin{equation}
{\rm L(\Delta E)}_i = 4\pi \left[{\rm D_A\left(1+z\right)}\right]^2
{\rm norm}_i\Lambda^{\rm xs}(T_i,Z_i; {\rm \Delta E}), \label{eqn.xspec}
\end{equation}
where we have used the symbol $Z_i$ to represent all the abundances,
$z$ is the redshift, $D_{\rm A}$ is the angular diameter distance
(cm), and $\Lambda^{\rm xs}$ is the plasma emissivity (erg s$^{-1}$
cm$^3$) corresponding to the {\sc xspec} implementation of the
relevant coronal plasma code (e.g., {\sc apec}, {\sc mekal}).  To give
a precise definition\footnote{In the {\sc xspec} Users' Manual,
norm$_i$ is called $K$. But the term ``norm'' is actually displayed
for this parameter for the coronal plasma models like {\sc apec}.} of
norm$_i$ that accounts for spatial variations of the temperature and
abundances within ${\rm V}_i$, we refer to the volume emissivity (erg
s$^{-1}$ cm$^{-3}$ ) of a coronal plasma,
\begin{equation}
\epsilon(\vec{x}; \Delta E) = n_e(\vec{x})n_{\rm H}(\vec{x})\Lambda(T(\vec{x}),Z(\vec{x});
{\rm\Delta E}), \label{eqn.volemis} 
\end{equation}
where, e.g., $T(\vec{x}) \equiv T(x,y,z)$, and the plasma emissivity
in {\sc xspec} is related to the conventional definition\footnote{This
expression is not given in the {\sc xspec} Users' Manual, but it
follows from the definition of norm$_i$ (i.e., $K$) in that manual.} 
by, $\Lambda^{\rm xs} \equiv 10^{14}\Lambda$. It is convenient to work
in terms of the volume mass density of the gas, $\rho_g(\vec{x})$,
rather than the volume number densities, $n_e(\vec{x})$ and
$n_H(\vec{x})$, separately. To a very good approximation, for this
calculation one may assume a fully ionized gas of pure H and He in
which case,
\begin{equation}
n_e = \frac{2+\mu}{5\mu}\frac{\rho_g}{m_u}, \hskip 1cm n_{\rm H} =
\frac{4-3\mu}{5\mu}\frac{\rho_g}{m_u},
\end{equation}
where $\mu$ is the mean atomic weight of the gas and $m_u$ is the
atomic mass unit. (Note that using $\mu=0.62$ corresponding to the
solar He abundance leads to $n_e=1.22n_{\rm H}$.) By setting equation
[\ref{eqn.xspec}] equal to the luminosity obtained by integrating
equation [\ref{eqn.volemis}] over the volume V$_i$ and solving for
norm$_i$ one finds,
\begin{equation}
{\rm norm}_i = \left[
\frac{10^{-14}}{4\pi[{\rm D_A(1+z)}]^2}\frac{(2+\mu)(4-3\mu)}{(5\mu)^2m_u^2}\right]
\frac{1}{\Lambda^{\rm xs}(T_i,Z_i;{\rm \Delta E})}
\int_{{\rm V}_i}\rho_g^2(\vec{x})\Lambda^{\rm
xs}(T(\vec{x}),Z(\vec{x});{\rm \Delta E}){{\rm d}^3x}, \label{eqn.norm}
\end{equation}
where ${\rm norm}_i \propto \int \rho_g^2{\rm d}^3x$ if $\Lambda^{\rm
xs}$ is constant over the volume $V_i$. If $\Lambda^{\rm xs}$ varies
over the volume, then the parameters $T_i$ and $Z_i$ obtained from
fitting a 1T coronal plasma model to the spectrum with multiple
temperatures and abundances will reflect average quantities weighted
by the emission profile within the volume and the detector
response.  We defer treatment of the detector response to the end of
this section and focus now on the deprojection and projection of
emission-weighted spherical quantities.

\bigskip

\noindent {\it Deprojection Analysis}: If the spectra are deprojected so that the
norm$_i$ values refer to spherical shells, then equation
[\ref{eqn.norm}] can be immediately recast in terms of the
weighted square of the gas density,
\begin{eqnarray}
\langle \rho_g^2\rangle_i & \equiv & \left[ \frac{4\pi[{\rm
D_A(1+z)}]^2}{10^{-14}}\frac{(5\mu)^2m_u^2}{(2+\mu)(4-3\mu)}\right]
\frac{{\rm norm}_i}{{\rm V}_i}  \\
& = & \frac{1}{\Lambda^{\rm xs}(T_i,Z_i;{\rm \Delta E})}\frac{3}{(r_i^3 - r_{i-1}^3)}
\int_{{\rm r}_{i-1}}^{{\rm r}_i}\rho_g^2(r)\Lambda^{\rm
xs}(T(r),Z(r);{\rm \Delta E}){r^2{\rm d}r}, \label{eqn.deproj}
\end{eqnarray}
where ${\rm V}_i= \frac{4\pi}{3}(r_i^3 - r_{i-1}^3)$ is the volume of
the spherical shell. Note, however, that many deprojection programs
like {\sc projct} in {\sc xspec} and others based on the widely used
``onion peeling'' deprojection method \citep{fabian81} give norm$_i$
with respect to the volume, $\frac{4\pi}{3}(r_i^2 - r_{i-1}^2)^{3/2}$,
representing the intersection of the 3D shell with the cylinder of the
same inner and outer radii (see equation \ref{eqn.vint} below). In
this case the integral in equation [\ref{eqn.deproj}] must proceed
over the intersecting volume and therefore depends not only on $r$.
Typically we perform the radial integrations by dividing up the shells
into 5-10 sub-shells each of constant $\rho_g$ and $\Lambda^{\rm xs}$,
where the volume of each sub-shell $j$ is $\sum_{k=m(i-1)+1}^{j}V^{\rm
int}_{kj}$ using the definitions associated with equations
[\ref{eqn.shellint}] and [\ref{eqn.vint}] in the projection analysis
below. The quantity $\sqrt{\langle \rho_g^2\rangle_i}$ is a measure of
the average density within the relevant volume (intersecting or total)
of the spherical shell.

The average temperature within the shell $i$ is given by,
\begin{equation}
\langle T\rangle_i = 
\frac{\int_{{\rm r}_{i-1}}^{{\rm r}_i}T(r)\rho_g^2(r)\Lambda^{\rm
xs}(T(r),Z(r);{\rm \Delta E}){r^2{\rm d}r}}{\int_{{\rm r}_{i-1}}^{{\rm r}_i}\rho_g^2(r)\Lambda^{\rm
xs}(T(r),Z(r);{\rm \Delta E}){r^2{\rm d}r}} ,
\end{equation}
where, as above, the integration proceeds instead only over the
intersecting volume if necessary. We set $\langle T\rangle_i=T_i$,
which holds exactly for constant $\Lambda^{\rm xs}$ within the
shell. (Similarly, for any abundance, such as iron, we set $\langle
Z_{\rm Fe}\rangle_i=Z_{{\rm Fe},i}$.) Since $\Lambda^{\rm xs}$
generally varies monotonically with increasing radius within the
shell, this association becomes increasingly more accurate as the
shell width is allowed to decrease.  However, even if calibration
uncertainties and other issues associated with the detector response
can be ignored, systematic errors associated with, e.g., background
subtraction, assumption of spherical symmetry, Galactic absorption,
may bias the inferred 1T model fitted to the multi-component spectral
data over a limited energy range. It is therefore essential to examine
the sensitivity of one's analysis to such effects as we have done here
and previously \citep{lewi03a,buot04a,humphrey06a,zappacosta06}.

\bigskip

\noindent {\it Projection Analysis}: This is the primary method employed in this
paper. The 1T models are fitted directly to the spectra that are
extracted from concentric circular annuli on the sky. Rather than
dividing norm$_i$ by the emitting volume, now we divide it by the area
of the annulus, ${\rm A}_i=\pi (R_i - R_{i-1})^2$, to obtain a
line-of-sight, emission-weighted projection of $\rho_g^2$ averaged
over A$_i$,
\begin{eqnarray}
{\rm proj} \, \, \langle \rho_g^2 \rangle_i & \equiv & \left[ \frac{4\pi[{\rm
D_A(1+z)}]^2}{10^{-14}}\frac{(5\mu)^2m_u^2}{(2+\mu)(4-3\mu)}\right]\frac{1}{{\rm
A}_i} {\rm norm}_i \\
& = & \frac{1}{\Lambda^{\rm xs}(T_i,Z_i;{\rm \Delta E})}\frac{2}{(R_i^2 - R_{i-1}^2)}
\int_{{\rm R}_{i-1}}^{{\rm R}_i}R{\rm d}R\int_{\rm los}\rho_g^2(r)\Lambda^{\rm
xs}(T(r),Z(r);{\rm \Delta E}){{\rm d}z}, \label{eqn.proj}
\end{eqnarray}
where $r=\sqrt{R^2 + z^2}$ and the projection proceeds within a
cylinder denoted by $i$ defined by inner radius, $R_{i-1}$, outer
radius, $R_i$, and height specified by the l.o.s.\
integration. (Typically we set the l.o.s.\ integration limits to
$\approx \pm 1.5$ virial radii, though our results are very
insensitive to this choice for values greater than the virial radius.)
For the special case of constant $\Lambda^{\rm xs}$ within the
cylinder $i$ we have that ${\rm proj} \, \, \langle \rho_g^2
\rangle_i$ equals $\int_{\rm los}\rho^2_gdz$ averaged over ${\rm A}_i$
\citep[e.g.,][]{buot04a}. 

In practice, for fast numerical evaluation it is preferable to
approximate the integrations in equation [\ref{eqn.proj}] in terms of
the contributions from discrete shells \citep{kriss83}.  The spherical
volume of the cluster is partitioned into a series of $N$ concentric
spherical shells such that, $r_0<r_1<r_2<\dots < r_N$, where the
number of shells and their widths are chosen to achieve desired
computational accuracy. A particular shell, $[r_{k-1},r_k]$, is
denoted by the index $k$ of the outer radius.  We define a
corresponding set of $N$ concentric circular annuli on the sky such
that, $R_0=r_0<R_1=r_1<\dots < R_N=r_N$, where the origins of the
annuli and shells coincide. In analogy to the 3D shells, we denote a
particular annulus, $[R_{k-1},R_k]$, by the index $k$ of the outer
radius. To insure computational accuracy this set of reference annuli
necessarily over-samples the set of annuli used to extract the X-ray
data. Consequently, for each annulus $i$ of the data, $[R_{i-1},R_i]$,
one defines a mapping between $i$ of the data annuli and $k$ of the
reference annuli so that the annulus $i$ contains multiple reference
annuli. It is useful to consider the case were the reference annuli
over-sample the data by some integer $m$ so that $k=mi$; i.e., annulus
$[R_{i-1},R_i]$ of the data contains all reference annuli between
$R_{m(i-1)}$ and $R_{mi}$\footnote{Since the data do not extend out to
the adopted edge of the system, it follows that $r_N\gg r_{mD}$, where
$D$ is the number of data annuli.}.  In this case equation
[\ref{eqn.proj}] may be approximated as,
\begin{equation}
{\rm proj} \, \, \langle \rho_g^2 \rangle_i  \simeq   \frac{1}{\Lambda^{\rm xs}(T_i,Z_i;{\rm \Delta
E})}\frac{1}{\pi(R_i^2 - R_{i-1}^2)}\sum_{j=m(i-1)+1}^{mi}\sum_{k=j}^N\left(\rho_g^2\Lambda^{\rm
xs}\right)_k{\rm V}^{\rm int}_{kj}, \label{eqn.shellint} 
\end{equation}
The inner sum $\sum_{k=j}^N$ projects
$\left(\rho_g^2\Lambda^{\rm xs}\right)_k
\equiv \rho_g^2(\bar{r}_k)\Lambda^{\rm
xs}(T(\bar{r}_k),Z(\bar{r}_k);\Delta {\rm E})$ into reference annulus
$j$, where $\bar{r}_k$ represents an intermediate radius within shell
$k$. The projection is carried out via the matrix,
\begin{equation}
{\rm V}^{\rm int}_{kj} =  \frac{4\pi}{3}\left[
\left(r_k^2 - R_{j-1}^2\right)^{\frac{3}{2}} - \left(r_k^2 - R_j^2\right)^{\frac{3}{2}} +
\left(r_{k-1}^2 - R_j^2\right)^{\frac{3}{2}} - \left(r_{k-1}^2 -
R_{j-1}^2\right)^{\frac{3}{2}}\right], \hskip 1cm k\ge j, \label{eqn.vint}
\end{equation}
representing the volume of shell $k$ that intersects the cylinder
defined by the width of annulus $j$ and infinite height
\citep{kriss83}.  If any terms in equation [\ref{eqn.vint}] have
negative arguments they must be set to zero; e.g., if $k=j$ then ${\rm
V}^{\rm int}_{kk} = \frac{4\pi}{3}(r_k^2 - r_{k-1}^2)^{3/2}$, as noted
above in the deprojection analysis. 

The average temperature within annulus $i$ is given by,
\begin{eqnarray}
\langle T\rangle_i & = &\frac{\int_{{\rm R}_{i-1}}^{{\rm R}_i}R{\rm d}R\int_{\rm
los}T(r)\rho_g^2(r)\Lambda^{\rm xs}(T(r),Z(r);{\rm \Delta
E}){{\rm d}z}}{\int_{{\rm R}_{i-1}}^{{\rm R}_i}R{\rm d}R\int_{\rm
los}\rho_g^2(r)\Lambda^{\rm xs}(T(r),Z(r);{\rm \Delta E}){{\rm
d}z}}\\
& \simeq & \frac{\sum_{j=m(i-1)+1}^{mi}\sum_{k=j}^N\left(T\rho_g^2\Lambda^{\rm xs}\right)_k
{\rm V}^{\rm int}_{kj}}{\sum_{j=m(i-1)+1}^{mi}\sum_{k=j}^N
\left(\rho_g^2\Lambda^{\rm xs}\right)_k{\rm V}^{\rm int}_{kj}}, \label{eqn.tempemis}
\end{eqnarray}
where again we set $\langle T\rangle_i=T_i$ with the same caveats
noted above in the deprojection analysis.

\bigskip

\noindent {\it Response Weighting}:
The existence of radial gradients in the temperature and abundances of
galaxy groups and clusters imply that the X-ray spectra evaluated over
spherical shells or circular annuli of finite width are not 1T coronal
plasmas. In theory, fitting a 1T model to such a multi-component
plasma yields average values of the spectral parameters norm$_i$,
$T_i$, and $Z_i$ suitable for comparison with the emission weighted
spherical models described previously.  Unfortunately, biased
parameter values can result from such 1T fits. Fitting a 1T coronal
plasma model to a multi-temperature spectrum with average temperature
near 1~keV and solar abundances results in a severe underestimate of
the iron abundance \citep[``Fe Bias'', see
][]{buot98c,buot00a,buot00c}.  But since the inferred temperature is
not biased \citep{buot00a}, and the abundance underestimate does not
translate to a large overestimate of the gas density, the effect on
the derived mass profile is minimal.  Note that the Fe Bias primarily
originates from the differences between the strengths of the Fe~L
shell lines in single and multi-temperature plasmas -- not from
effects related to the detector response other than restricted
bandwidth.

However, since the effective area of the detector responses of ACIS on
\chandra\ and EPIC on \xmm\ (as well as detectors on previous X-ray
satellites like {\sl ROSAT}) peak near 1~keV, fitting a 1T model to a
spectrum having a range of temperature components
above 1~keV will tend to yield a temperature that is biased low
\citep{mazzotta04,vikh06b}.  This effect is most pronounced for
the ``projection analysis'' of galaxy clusters possessing in each
annulus a wide range of temperatures above $\sim 3$~keV where the Fe~L
shell lines are weak. (At lower temperatures, the Fe Bias applies as
noted above.) Approximate methods to account for this effect have been
proposed by \citet{mazzotta04} and \citet{vikh06b}.

Our more rigorous approach is a straightforward averaging of the
plasma emissivity over the detector response for the (projected)
region in question. Let the response matrix that determines the
probability a photon of energy $E$ will be detected in pulse height
analyzer (PHA) bin $n$ be denoted by ${\rm RSP}_i(n,E)$, where $i$
denotes the annulus on the sky as defined in the ``projection
analysis'' above. This response matrix is usually considered to be the
product of a ``redistribution matrix'', ${\rm RMF}_i(n,E)$, and an
``auxiliary response file'', ${\rm ARF}_i(E)$, the latter of which
contains the information on the effective area. Then the count rate in
PHA bin $n$ is proportional to $\rho_g^2\sum_{E}\Lambda^{\rm xs}(T,Z;
E){\rm RSP}_i(n,E)$, where the sum is over all energies in the
response matrix. By summing over all PHA bins $n$ corresponding to the
energy range $[E_1,E_2]$, and since $\rho_g^2$ does not depend on the
convolution, we may account for the detector response in our above
presentation by replacing $\Lambda^{\rm xs}$ with,
\begin{equation}
\overline{\left(\Lambda^{\rm xs}(T,Z;\Delta \rm E)\right)}_i \equiv
\sum_n\sum_E\Lambda^{\rm xs}(T,Z;E){\rm RSP}_i(n,E), \label{eqn.rsp}
\end{equation}
where we have not renormalized since the normalization of $\Lambda^{\rm
xs}$ does not need to be specified for ${\rm proj} \, \, \langle
\rho_g^2 \rangle_i$ and $\langle T\rangle_i$.

Although applying this response weighting helps to mitigate the
temperature bias for hot systems, this procedure is still not formally
equivalent to that used to obtain $T_i$ and $Z_i$ and norm$_i$ by
fitting a 1T model to a spectrum containing multiple temperature and
abundance components.  To avoid the Fe Bias and response bias
altogether, the projected spherical models must be fitted directly to
the observed spectra. This means rather than predicting quantities
that are integrated over the bandpass, one must predict the photon
count rate C$_{i,n}$ in circular annulus $i$ for each PHA bin $n$,
\begin{eqnarray}
{\rm C}_{i,n} & = &  \left[ \frac{10^{-14}}{2[{\rm
D_A(1+z)}]^2}\frac{(2+\mu)(4-3\mu)}{(5\mu)^2m_u^2}\right]
\int_{{\rm R}_{i-1}}^{{\rm R}_i}R{\rm d}R\int_{\rm los}{{\rm
d}z}\rho_g^2(r)\sum_E\Lambda^{\rm xs}(T(r),Z(r);E){\rm
RSP}_i(n,E) \nonumber \\
& \simeq &  
\left[ \frac{10^{-14}}{4\pi[{\rm
D_A(1+z)}]^2}\frac{(2+\mu)(4-3\mu)}{(5\mu)^2m_u^2}\right]
\sum_{j=m(i-1)+1}^{mi}\sum_{k=j}^N{\rm V}^{\rm int}_{kj} 
\left(\rho_g^2\right)_k \sum_E\left(\Lambda^{\rm xs}(E)\right)_k{\rm
RSP}_i(n,E),  
\end{eqnarray}
such as done, essentially, in a procedure like SMAUG
\citep{pizz03a}. Note that $(\rho_g^2)_k\equiv \rho_g^2(\bar{r}_k)$
and $(\Lambda^{\rm xs}(E))_k\equiv \Lambda^{\rm
xs}(T(\bar{r}_k),Z(\bar{r}_k); E)$ with $\bar{r}_k$ an intermediate
radius in shell $k$ as above.

It is our experience that even for systems possessing the highest
quality data available from \chandra\ and \xmm, the magnitudes of
statistical errors and the key systematic errors noted above are
sufficiently large so that presently there is little advantage to
increasing the sophistication of the model comparison beyond that
expressed by equation [\ref{eqn.rsp}].  (Note also that we do not use
SMAUG in this paper because its current implementation in {\sc xspec}
does not include the ``parametric mass'' fitting approach in \S
\ref{sect_mass}.)  However, it is expected that data from the next
generation of X-ray satellites (Con-X, Xeus) will be of sufficiently
high quality to require the direct fitting approach for large numbers
of systems, provided there is strict control of other systematic
errors.

\bibliographystyle{apj}
\bibliography{dabrefs,fgrefs}

\end{document}